\documentclass[11pt]{article}
\usepackage{fullpage}
\usepackage{times}
\usepackage{algorithmic}
\usepackage[ruled]{algorithm2e}
\usepackage{subfigure}
\usepackage{color}
\usepackage{url}
\usepackage{graphicx}
\usepackage{amsmath}
\usepackage{amssymb}

\usepackage{tikz}
\usetikzlibrary{plotmarks}
\usetikzlibrary{arrows,decorations.pathmorphing}
\usetikzlibrary{decorations.pathreplacing}

\tikzset{
    node style sp/.style={draw,circle,minimum size=\myunit},
    node style ge/.style={circle,minimum size=\myunit},
    arrow style mul/.style={draw,sloped,midway,fill=white},
    arrow style plus/.style={midway,sloped,fill=white},
}

\newcommand{\qed}{\mbox{}\hspace*{\fill}\nolinebreak\mbox{$\rule{0.6em}{0.6em}$}
}
\newcommand{\expect}{{\bf \mbox{\bf E}}}
\newcommand{\prob}{{\bf \mbox{\bf Pr}}}

\definecolor{gray}{rgb}{0.5,0.5,0.5}
\newcommand{\e}{{\epsilon}}

\newcommand{\get}{\text{TREE-COMBINE}}

\newtheorem{theorem}{Theorem}
\newtheorem{lemma}[theorem]{Lemma}
\newtheorem{claim}[theorem]{Claim}
\newtheorem{fact}[theorem]{Fact}
\newtheorem{corollary}[theorem]{Corollary}
\newtheorem{definition}[theorem]{Definition}
\newtheorem{remark}[theorem]{Remark}

\newenvironment{proof}{{\bf Proof:}}{$\qed$\par}

\newenvironment{proofof}[1]{\noindent{\bf Proof of #1:}}{$\qed$\par}

\usepackage{hyperref}
\hypersetup{
bookmarksnumbered
}

\newcommand{\erf}{\text{erf}}

\begin{document}

\title{\Large Prediction strategies without loss\thanks{A preliminary version of this paper appeared in \cite{mab-nips}.}}
\setcounter{page}{0}
\author{
Michael Kapralov\thanks{Institute for Computational and Mathematical Engineering, Stanford University, Stanford, CA 94305. 
    Email:~{\tt kapralov@stanford.edu}. Part of this work was done when the author was an intern at Microsoft Research Silicon~Valley.}
\and Rina Panigrahy\thanks{Microsoft Research Silicon Valley, Mountain View, CA 94043. 
    Email:~{\tt rina@microsoft.com}}}

\maketitle

\begin{abstract}
Consider a sequence of bits where we are trying to predict the next bit from the previous bits. Assume we are allowed to say `predict 0' or `predict 1', and our payoff is $+1$ if the prediction is correct and $-1$ otherwise. We will say that at each point in time the loss of an algorithm is the number of wrong predictions minus the number of right predictions so far.  In this paper we are interested in algorithms that have essentially zero (expected) loss over any string at any point in time and yet have small regret with respect to always predicting $0$ or always predicting $1$. For a sequence of length $T$ our algorithm has  regret $14\epsilon T $ and loss   $2\sqrt{T}e^{-\epsilon^2 T} $ in expectation for all strings. We show that the tradeoff between loss and regret is optimal up to constant factors.

Our techniques extend to the general setting of $N$ experts, where the related problem of trading off regret to the best expert for regret to the `special' expert has been studied by Even-Dar et al. (COLT'07). We obtain essentially zero loss with respect to the special expert and optimal loss/regret tradeoff,
improving upon the results of Even-Dar et al and settling the main question left open in their paper.

 The strong loss bounds of the algorithm have some surprising consequences. A simple iterative application of our algorithm gives essentially optimal regret bounds at multiple time scales, bounds with respect to $k$-shifting optima as well as regret bounds with respect to higher norms of the input sequence.
 
\end{abstract}


\section{Introduction}

Consider a gambler who is trying to predict the next bit in a sequence of bits.  One could think of the bits as indications of whether a stock price goes up or down on a given day, where we assume that the stock always goes up or down by $1$  (this is, of course, a very simplified model of the stock market). If the gambler predicts $1$ (i.e. that the stock will go up), she buys one stock to sell it the next day, and short sells one stock if her prediction is $0$. We will also allow the gambler to bet fractionally by letting him specify a confidence $c$ where $0 \le c \le 1$ in his prediction. If the prediction is right the gambler gets a payoff of $c$ otherwise $-c$.  While the gambler is tempted to make predictions with the prospect of making money, there is also the risk of  ending up with a loss.  Is there a way to never end up with a loss?   Clearly there is the strategy of never predicting (by setting confidence $0$) all the time that never has a loss but also never has a positive payoff.  However, if the sequence is very imbalanced and has many more  $0$'s than $1$'s then this never predict strategy has a high regret with respect to the strategy that predicts the majority bit.  Thus, one is interested in a strategy that has a small regret with respect to predicting the majority bit and incurs no loss at the same time.

 Our main result is that while one cannot always avoid a loss and still have a small regret,  this is possible if we allow for an {\em exponentially small} loss.  More precisely, we show that for any $\e>1/\sqrt{T}$ there exists an algorithm that achieves regret at most $14\e T$ and  loss at most $2e^{-\e^2T}\sqrt{T}$, where $T$ is the time horizon. Thus, the loss is {\em exponentially small in the length of the sequence}.  
 
 The bit prediction problem can be cast as the experts problem with two experts: $S_+$, that always predicts $1$ and $S_-$ that always predicts $0$. This problem has been studied extensively, and very efficient algorithms are known. The weighted majority algorithm of \cite{weighted-majority} is known to give optimal regret guarantees. However, it can be seen that weighted majority may result in a loss of $\Omega(\sqrt{T})$.  The best known result on bounding loss  is the work of Even-Dar et al. \cite{kearns-regret}
 on the problem of trading off regret to the best expert for regret to the average expert, which is equivalent to our problem. Stated as a result on bounding loss, they were able to obtain a constant loss and regret  $O(\sqrt{T}\log T)$. Their work  left the question open as to whether it is possible to even get a regret of  $O(\sqrt{T\log T})$ and constant loss. In this paper we give an optimal regret/loss tradeoff, in particular showing that this regret can be achieved even with subconstant loss. 
 
Our results extend to the general setting of  prediction with expert advice when there are multiple experts. In this problem the decision maker iteratively chooses among $N$ available alternatives without knowledge of their payoffs, and gets payoff based on the chosen alternative. The payoffs of all alternatives are revealed after the decision is made. This process is repeated over $T$ rounds, and the goal of the decision maker is to maximize her cumulative payoff over all time steps $t=1,\ldots, T$. This problem and  its variations has been studied extensively, and efficient algorithms have been obtained (e.g.~\cite{cover-binary, weighted-majority, cover-portfolios, auer-nonstoch, audibert-bubeck}). The most widely used measure of performance of an online decision making algorithm is {\em regret}, which is defined as the difference between the payoff of the best {\em fixed} alternative and the payoff of the algorithm. The well-known weighted majority algorithm of \cite{weighted-majority} obtains regret $O(\sqrt{T\log N})$ even when no assumptions are made on the process generating the payoff. Regret to the best fixed alternative in hindsight is a very natural notion when the payoffs are sampled from an unknown distribution, and in fact such scenarios show that the bound of $O(\sqrt{T\log N})$ on regret achieved by the weighted majority algorithm is optimal. 
 
 Even-Dar et al. \cite{kearns-regret} gave an algorithm that has {\em constant} regret to any fixed distribution on the experts at the expense of regret $O(\sqrt{T\log N}(\log T+\log\log N))$ with respect to all other experts\footnote{In fact, \cite{kearns-regret} provide several algorithms, of which the most relevant for comparison are {\em Phased Agression}, yielding  $O(\sqrt{T\log N}(\log T+\log\log N))$ regret to the best and  D-Prod, yielding $O(\sqrt{T/\log N}\log T)$ regret to the best. For the bit prediction problem one would set $N=2$ and use the uniform distribution over the `predict 0' and `predict 1' strategy as the special distribution. Our algorithm improves on both of them, yielding an optimal tradeoff.}.  We obtain an optimal  tradeoff between the two, getting an algorithm with regret $O(\sqrt{T(\log N+\log T)})$ to the best and $O((NT)^{-\Omega(1)})$ to the average as a special case. We also note, similarly to \cite{kearns-regret} that our regret/loss tradeoff cannot be obtained by using standard regret minimization algorithms with a prior that is concentrated on the `special' expert, since the prior would have to put a significant weight on the `special' expert, resulting in $\Omega(T)$ regret to the best expert.
  
  The extension to the case of $N$ experts uses the idea of improving one expert's predictions by that of another.
The strong loss bounds of our algorithm allow us to achieve {\em lossless boosting}, i.e. we use available expert to continuously improve upon the performance of the base expert whenever possible while essentially never hurting its performance. When comparing two experts, we track the difference in the payoffs discounted geometrically over time and apply a transform 
$g(x)$  on this difference to  obtain a weighting that is applied to give a linear combination of the two experts with a higher weight being applied on the expert with a higher discounted payoff. The shape of $g(x)$ is given by $\erf\left(\frac{x}{4\sqrt{T}}\right)e^{x^2/(16T)}$, capped at $\pm 1$ (we refer to our predictor as DISCOUNTED-NORMAL-PREDICTOR). The weighted majority algorithm on the other hand uses a transform with the shape of the $\tanh(\frac{x}{\sqrt{T}})$ function and ignores geometric discounting~(see~Figure~\ref{fig:1}).

  An important property of our algorithm is that it does not need a high imbalance between the number of ones and the number of zeros in the whole sequence to have a gain: it is sufficient for the imbalance to be large enough in {\em at least one contiguous time window}\footnote{More precisely, we use an infinite window with geometrically decreasing weighting, so that most of the weight is contained in the window of size $O(n)$, where $n$ is a parameter of the algorithm. }, the size of which is a parameter of the algorithm. This property allows us to easily obtain optimal {\em adaptive regret} bounds, i.e.  we show that the payoff of our algorithm in {\em any geometric window} of size $n$  is at most $O(\sqrt{n\log (NT)})$ worse than the payoff of the strategy that is {\em best in that window} (see  Theorem~\ref{thm:Z-rand}). 
  In section~\ref{sec:k-shifting} we also obtain bounds against the class of strategies that are allowed to change experts multiple times while maintaining the essentially zero loss property.   We note that even though similar bounds  (without the essentially zero loss property) have been obtained before (\cite{blum-mansour, freund-schapire-singer-warmuth,  vovk} and, more recently, \cite{hazan-seshadhri}), our approach is very different and arguably simpler. 

We are also able to obtain regret bounds that depend on the complexity of the bit sequence. One measure of complexity of a bit sequence its Kolmogorov complexity which is not computable. For a suitable time bounded variant of the Kolmogorov complexity we show that one can obtain regret and loss bounds that depend on the complexity of the string. 

 We also obtain bounds against $k$-shifting optima (Theorem~\ref{thm:windows-adaptive}) in a parameter-free fashion while maintaining the essentially zero loss property.  In this setting the $T$ time steps are partitioned into intervals and a different strategy may be used in each partition. In such a case our regret is at most $\sum_j O\left(\sqrt{|I_j| \log (1/Z)}\right)$, where $|I_j|$ is the length of the $j$-th interval. As well as achieving these regret bounds in a parameter-free fashion we also preserve the low loss property. 

 Additionally, we show how risk-free assets can be constructed using our algorithm under the assumption of bounded change of price of a stock. This application motivates studying the effect of transaction costs on the algorithm. It turns out that one can still get bounded loss at the expense of making regret commensurate with the transaction cost. This involves  treating the confidence values of the algorithm as probabilities of selling or buying. We derive the corresponding high probability bounds on the loss of the algorithm in section~\ref{app:risk-free}.
    
    Finally, we show that our techniques can be applied to the multi-armed bandit problem with partial information (see, e.g.~\cite{auer-nonstoch}), giving an algorithm with $O(N^{1/3} T^{2/3}\log^{1/3}(NT))$ regret and loss $O((NT)^{-2})$ with respect to the average of all arms. Additionally, we show how our framework can be applied to the online convex optimization algorithm of Zinkevich \cite{zinkevich} to obtain an algorithm with good adaptive regret guarantees (see section~\ref{app:app}). 

We would like to point out that our work on this problem was motivated from observing a psychological tendency in prediction: if one has recently seen a large number of 0's, there is a strong psychological inclination to predict a 0 for the next bit.  We were interested in finding if this instinct of being strongly influenced by recent bits is misleading or if it has some rational basis.  Our results show that in certain settings this instinct indeed has a rational basis and gives a small loss. In fact it can be shown that if one does not discount the older bits and only predicts based on the difference  in the number of $0$'s and $1$'s seen so far then it is impossible to achieve a small loss (see claim~\ref{nodiscount}.)

\subsection{Related work}

In the general online decision problem the decision maker has to choose  a decision from a set of available alternatives at each point in time $t=1,\ldots, T$ without knowing future payoffs of the available alternatives. At each time step $t$ the payoffs of alternatives at time $t$ are revealed to the decision maker after she commits to a choice. Online decision problems have been studied under different feedback models and assumptions on the process generating the payoffs. The transparent feedback, or full information, model costs of all available alternatives are revealed, while in the opaque feedback, or partial information model only the cost of the decision that was made is revealed to the algorithm. 
The performance of an online decision making algorithm is usually measured in terms of regret, i.e. the difference between the payoff of the algorithm and the payoff of the best fixed alternative in hindsight.

Various assumptions on the process generating the payoff of arms have been considered in the literature. When a prior belief on the distribution of payoffs is assumed, the discounted reward with infinite time horizon can be efficiently maximized using the Gittins index (see, e.g. \cite{gittins, tsitsiklis}). Low-regret algorithms for the setting when the payoffs come from an unknown probability distribution were obtained in \cite{auer, agarwal, lai-robbins}. Assumptions on the payoff sequence are not necessary to achieve low regret. In particular, the well-known weighted majority algorithm \cite{weighted-majority} yields $O(\sqrt{T\log N})$ regret in the full information model (also known as the experts problem).  Surprisingly, \cite{auer-nonstoch} showed that low regret with respect to the best arm in hindsight can be achieved without making any assumptions on the payoff sequence even in the partial information model, giving the first algorithm with $O(\sqrt{NT\log N})$ regret in this setting.
 Better bounds have been obtained under the assumption that the sequence of payoffs has low variance (e.g. \cite{hazan-kale}). A related line of work applying similar techniques to problems in finance includes \cite{cover-portfolios, kalai-vempala, hazan-kale-invest}.

More specialized techniques have been developed for the online optimization problem  in both the full and partial information models (\cite{kalai-vempala-jcss, dani-hayes, awerbuch-kleinberg}). Better bounds can be obtained under the convexity assumption (\cite{zinkevich, no-gradient, dani-hayes, log-online-convex}). 
Another line of work focuses on obtaining good regret guarantees when the space of available alternatives is very large or possibly infinite, but has some special structure (e.g. forms a metric space) -- \cite{bandits-metric, dichotomies}.  It is hard to faithfully represent the large body of work on online decision problems in limited space, and we refer the reader to \cite{plg} for a detailed exposition. 

Other measures of performance of an online algorithm have been considered in the literature. The question of which tradeoff between can be achieved if one would like to have a significantly better guarantee with respect to a fixed arm or a distribution on arms was asked before in \cite{kearns-regret} as we discussed in the introduction.  Besides improving on the result of \cite{kearns-regret}, we also answer the question left open by the authors: `It is currently unknown whether or not it is possible to strengthen Theorem~6 to say that any algorithm with regret $O(\sqrt{T\log T})$ to the best expert must have regret $\Omega(T^\e)$ to the average for some constant $\e>0$'. In fact, for any $\gamma>0$ our algorithm has loss $O(T^{-\gamma})$ (corresponding to regret to the average) when the regret is $O(\sqrt{\gamma T\log T})$, thus showing that such a strengthening is impossible. Tradeoffs between regret and loss were also examined in \cite{vovk-game}, where the author studied the set of values of $a, b$ for which an algorithm can have payoff $a OPT+b\log N$, where $OPT$ is the payoff of the best arm and $a, b$ are constants.  The problem of bit prediction was also considered in \cite{freund-coin}, where several loss functions are considered. None of them, however, corresponds to our setting, making the results incomparable. 

In recent work on the NormalHedge algorithm\cite{normalhedge} the authors use a potential function which is very similar to our function $g(x)$ (see \eqref{eq:def-g} below), getting strong regret guarantees to the $\e$-quantile of best experts. However, the use of the function $g(x)$ seems to be quite different from ours, as is the focus of the paper \cite{normalhedge}.

\subsection{Preliminaries}
We start by defining the bit prediction problem formally. Let $b_t,t=1,\ldots, T$ be an adversarial sequence of bits. It will be convenient to adopt the convention that $b_t\in \{-1,+1\}$ instead of $b_t\in \{0, 1\}$ since it simplifies the formula for the payoff. In fact, in what follows we will only assume that  $-1\leq b_t\leq 1$, allowing $b_t$ to be real numbers.  At each time step $t=1,\ldots, T$ the algorithm is required to output a {\em confidence level} $f_t\in [-1, 1]$, and then the  value of $b_t$ is revealed to it. The payoff of the algorithm by time $t'$ is 
\begin{equation}
A_{t'}=\sum_{t=1}^{t'} f_t b_t.
\end{equation}
 For example, if $b_t\in \{-1, +1\}$, then this setup is analogous to a prediction process in which a player observes a sequence of bits and at each point in time predicts that the value of the next bit will be $\text{sign}(f_t)$ with confidence $|f_t|$.  Predicting  $f_t\equiv 0$ amounts to not playing the game, and incurs no loss, while not bringing any profit. We define the loss of the algorithm on a string $b$ as
\begin{equation*}
\text{loss}=\min\{-A_t, 0\},
\end{equation*}
i.e. the absolute value of the smallest negative payoff over all time steps.

 It is easy to see that any algorithm that has a positive expected payoff on some sequence necessarily loses on another sequence. Thus, we are concerned with finding a prediction strategy  that has {\em exponentially small loss bounds} but also has {\em low regret} against a number of given prediction strategies.  In the simplest setting we would like to design an algorithm that has low regret against two basic strategies: $S_{+}$, which always predicts $+1$ and $S_-$, which always predicts $-1$. Note that the maximum of the payoffs of $S_+$ and $S_-$ is always equal to $\left|\sum_{t=1}^T b_t\right|$. We denote the base random strategy, which predicts with confidence $0$, by $S_0$. In what follows we will use the notation $A_T$ for the cumulative payoff of the algorithm by time $T$ as defined above. We will also use the notation $A_T(S)$ to denote the payoff of a strategy $S$ over $T$ time steps.
 
Note that the simplest setting just described in fact corresponds to the experts problem with two experts $S_+$ and $S_-$, where we are interested in designing an algorithm that has low regret with respect to the best of $S_+$ and $S_-$, at the same time having a small loss with respect to the {\em average} $S_0=\frac1{2}(S_++S_-)$. This provides the connection to the setting of \cite{kearns-regret}.

As we will show in section~\ref{sec:combine}, our techniques extend easily to give an algorithm that has low regret with respect to the best of any $N$ bit prediction  strategies and exponentially small loss. Our techniques work for the general experts  problem, where loss corresponds to regret with respect to the `special' expert $S_0$, and hence we give the proof in this setting. This provides the connection to the work of \cite{kearns-regret}.
 
 In section~\ref{sec:main} we give an algorithm for the case of two prediction strategies $S_+$ and $S_-$, and in section~\ref{sec:combine} we extend it  to the general experts problem.

\section{Results}

All results in the paper are based on the following main result:

\begin{theorem}\label{thm:main}
For any $\e\geq \sqrt{\frac{1}{T}}$ there exists an algorithm $A$ for which 
\begin{equation*}
A_T\geq \max\left\{\left|\sum_{j=1}^T b_j\right|-14\e T, 0\right\}-2\sqrt{T}e^{-\e^2 T},
\end{equation*}
i.e. the algorithm has at most $14\e T$ regret against $S_+$ and $S_-$ {\em as well as a exponentially small loss}.
By setting $\e$ so that the loss bound is $2Z\sqrt{T}$, we get a regret bound of $\sqrt{T\log(1/Z)}$. 
\end{theorem}

We note that the algorithm is a strict generalization of weighted majority, which can be seen by letting $Z=\Theta(1)$ (later we ill see that this property will also hold for the generalization to $N$ experts.)

The tradeoff between the loss and regret is optimal up to constant factors in the regret term:
\begin{theorem} \label{thm:lower-bound}
Any algorithm that has  regret  $O(\sqrt{T\log(1/Z)})$ incurs loss $\Omega(Z\sqrt{T})$ on at least one sequence of bits $b_t,t=1,\ldots,T$. 
\end{theorem}

We now state the generalization to $N$ strategies (experts).  Given $N$ prediction strategies (experts) $S_1, S_2, \cdots, S_N$, let $s_{i,t} \in [-1,1]$ denote the payoff of $S_i$ at time $t$. 
Let $S_{i,t}$ denote the cumulative payoff $\sum_{j=1}^t s_{i,j}$ of $S_j$ upto time $t$. For a prediction algorithm $A$, define $Regret(A) = \max_{i=1}^N S_{i,T} - A_T$. Note that earlier we were only considering regret with respect to $S_+$ and $S_-$. Loss is also defined as before. Loss with respect to a base strategy $S_0$ is defined as $\min\{-(A_t-S_{0,t}), 0\}$,

\begin{theorem}\label{thm:n-exp}
For any $Z<1/e$ there exists an algorithm for combining $N$ strategies that has regret $O(\sqrt{T\log (N/Z)})$ 
 against the best of $N$ strategies and loss at most $O(Z\sqrt{T})$ with respect to any strategy $S_0$ fixed a priori. These bounds are optimal up to constant factors. 
\end{theorem}

\subsection{\bf Uniform regrets at differerent time scales}

We are also able to derive  uniform  regret bounds with respect to $N$ strategies $S_1,\cdots, S_N$ at different time scales. We essentially show that if thre are $N$ strategies, then for any  window length $n$, our time discounted regret is $O(\sqrt {n\log(1/Z)})$; the regret will be measured not exactly over a window of length $n$ but will use geometrically decreasing weights adding upto $n$ instead of uniformly weighting the last $n$ time steps as defined next.

 First, for a sequence $w_t, t=1,\ldots, T$ of real numbers and for a parameter $\rho=1-1/n\in(0, 1)$ define 
\begin{equation*}
\tilde w^\rho_t=\sum_{j=1}^{t}\rho^{t-j} w_j.
\end{equation*}
Thus $\tilde w^\rho_t$ like the sum of the last $n$ values of $w_j$, except that instead of using strictly the last $n$ values, we use  geometrically decaying weights that adds up to $n$.

\begin{definition}
A sequence $w_j$ is \emph{$Z$-uniform at scale $\rho=1-1/n$} if  one has $\tilde{w_t}^\rho\leq c\sqrt{n\log (1/Z)}$ for all $1\leq t\leq T$, for some constant $c>0$.
\end{definition}

Note that if the input sequence is iid $\textbf{Ber}(\pm 1, 1/2)$, then it is $Z$-uniform at any scale with probability at least $1-Z$ for any $Z>0$.
We now prove that the difference between the payoff of our algorithm and the payoff of any expert is $Z$-uniform, i.e. does not exceed the standard deviation of a uniformly random variable in any sufficiently large window, when the loss is bounded by $Z$. More precisely,

\begin{theorem}\label{thm:Z-rand}
There is a prediction algorithm with payoff sequence $s^{*}_{t}$ such that the sequences $s_{j, t}-s^{*}_{t}$ are $Z$-uniform for any $1\leq j\leq N$ at any scale $\rho\geq 1-1/(80\log (1/Z))$ when $Z=o((NT)^{-2})$. Moreover, the loss of the algorithm with respect to the base strategy is at most $Z\sqrt{T}$.
\end{theorem}

Note that if $r_{j,t} = s_{j,t} - s^{*}_{t}$, then $\widetilde{r_{j,t}}^\rho$ is like the regret of our algorithm with respect to $S_j$ in the window of last $n$ steps, except that instead of looking exactly at the last $n$ steps we are looking at a geometrically decaying window with total weight $n$.

\subsection{Regret bounds based on complexity of bit sequence}

We will now show how to obtain a prediction algorithm whose payoff depends on the complexity of the string. One measure of the complexity of a string is its Kolmogorov Complexity which is the size of the smallest Turing machine that outputs that string. The Kolmogorov Complexity however is not computable. We could use a variant of this measure that looks at the the smallest Turing machine  that predicts the string in a certain bounded amount of  time complexity $T(n)$  (where $n$ may be the size of the string.)  Let $comp(S)$ to be size of a (possibly probabilistic) Turing machine $S$ and let $A_T(S)$ denote the payoff it achieves according to its predictions within the bounded time complexity $T(n)$. Then there is an (inefficient) algorithm with running time exponential in $n$ such that:

\begin{theorem}\label{thm:kol}
For any sequence of bits there is a prediction algorithm that achieves payoff at least
$$A_T \geq \max_{S} \left\{A_T(S)-O\left(\sqrt{T\text{comp}(S)}\right), 0\right\}-O\left(T^{-\Omega(1)}\right).
$$
\end{theorem}

We also obtain bounds against $k$-shifting optima (section \ref{sec:k-shifting}) in a parameter-free fashion while maintaining the essentially zero loss property.  In particular, we prove 
\begin{theorem}\label{thm:windows-adaptive}
Let $\{S_1,\ldots, S_N\}$ be a set of strategies and let $S_0$ be  a base strategy. There exists an algorithm for combining $S_0,\ldots, S_N$ whose payoff $A_T$ after $T$ time steps satisfies the following conditions. Let $I_1,\ldots, I_k, I_j=[A_j, B_j]\subset[1..T]$ be a covering of $[1..T]$ by disjoint intervals. Then for any assignment of strategies to intervals $\eta_j, 1\leq j\leq k$ one has
\begin{equation*}
\begin{split}
A_T&\geq\max\left\{\sum_{j=1}^k\left[S_{\eta_j}(I_j)-O\left(\sqrt{|I_{j}| \log (1/Z)}\right)\right],\text A_T(S_0)\right\}-O(ZNT\log T),
\end{split}
\end{equation*}
where $S_{\eta_j}(I)$ is the cumulative payoff of strategy $S_{\eta_j}$ on interval $I$.
\end{theorem}
Note that the guarantees on the payoff hold {\em for any partitioning} of the interval $[1..T]$ into subintervals, as stated above, and that the zero loss property with respect to $S_0$ is preserved.

Regret bounds based that depend on the $l_p$ norm of the costs are provided in section~\ref{sec:k-shifting} as well.  Finally in section~\ref{app:risk-free} we show how risk free assets may be constructed using our algorithms. In section~\ref{app:app} we show applications of our framework to multi-armed bandits with partial information and online convex optimization. 

\section{Main algorithm}\label{sec:main}

In this section we will prove the main Theorem~\ref{thm:main}.
We start by giving intuition about the algorithm by considering some natural approaches to bit prediction. What would one do to predict the next bit given the previous $n$ bits (Figure~\ref{fig:01}, left panel). Observe that one natural tendency is to predict based on the frequency of $0$'s and $1$'s  in the past. If all $n$ bits are $1$, should be predict $1$ with a high confidence?  What if they aren't all $1$ but there are many more $1$'s than $0$'s? One approach is to let the prediction confidence depend on the imbalance $x$($=$ number of $1$'s - number of $0$'s) in the bits seen so far. It is also natural to give more weight to the recent bits; for example consider the sequence in which there are more $0$'s but the last few bits are $1$.
Figure~\ref{fig:01}, right panel, shows some possible confidence functions for predicting $1$ based on $x$ (for example, the weighted majority uses the $\tanh(x/\sqrt{T})$ function). We will devise a function that allows one to bound the loss. We will weight the recent bits higher. The $i$-th last bit will have weight $(1-1/n)^i$; Thus, in order to predict $b_{t+1}$ from the previous bits, we compute the discounted deviation $x =\sum_{j=1}^{t-1} \rho^{t-1-j} b_j$. Our confidence function $g(x)$ will be essentially zero until $x > \tilde \Omega(\sqrt n)$ after which it shoots to $100$\% very fast. We will show that with this confidence function we will never incur a significant loss.


\begin{figure}
\begin{minipage}[b]{8 cm}
\begin{center}
$$
\begin{array}{c}
\text{...............00000000000000000000000\fbox{?}}\\
\\
\text{...............00000001111111111111111\fbox{?}}\\
\\
\text{...............00000010000010000100000\fbox{?}}\\
\\
\text{...............00000000000000000000111\fbox{?}}\\
\end{array}
$$
\end{center}
\vspace{0.7in}
\end{minipage}
\begin{minipage}[b]{5 cm}
\begin{center}
\begin{tikzpicture}[y=.8cm, x=0.8cm,font=\sffamily]

	\draw[->] (0,0) -- coordinate (x axis mid) (5,0);
    	\draw[->] (0,0) -- coordinate (y axis mid) (0,5);
	\draw (0,0) -- coordinate (x axis mid) (-5,0);
    	\draw (0,0) -- coordinate (y axis mid) (0,-5);
	
	\draw [semithick]  
   (-5,-5) to [out=45,in=45] (0,0) 
   to [out=45,in=45] (5,5);
   
   	\draw [semithick, dotted]  
   (-5,-5) to [out=0,in=-120] (0,0) 
   to [out=60,in=180] (5,5);

   	\draw [semithick, dashed]  
   (-5,-5) to [out=0,in=-180] (0,0) 
   to [out=0,in=180] (5,5);

         \draw(3, -1) node {deviation=\#1-\#0};

         \draw(4.5, +0.5) node {$+n$};
          \draw(-4.5, +0.5) node {$-n$};

         \draw(0.5, +4.5) node {$+1$};
          \draw(-0.5, -4.5) node {$-1$};

           \draw(-0.5, 2.5) node[rotate=90] {\small confidence of betting $1$};
           \draw(0.5, -2.5) node[rotate=90] {\small confidence of betting $0$};

\end{tikzpicture}
\end{center}
\end{minipage}
\caption{(Left) How should one predict the next bit? How should one weigh the recent bits vs the older bits? (Right) How confident should be in your prediction be based on the deviation in the number of $0$'s and $1$'s in the last $n$ bits? Which of these confidence functions is best?}
\label{fig:01}
\end{figure}

We note that the algorithm is a strict generalization of weighted majority, which can be seen by letting $Z=\Theta(1)$ (this property will also hold for the generalization to $N$ experts in section~\ref{sec:combine}). 

Our algorithm will have the following form. For a chosen discount factor $\rho=1-1/n, 0\leq \rho\leq 1$ the algorithm maintains a discounted deviation $x_t=\sum_{j=1}^{t-1} \rho^{t-1-j} b_j$ at each time $t=1,\ldots, T$. The value of the prediction at time $t$ is then given by $g(x_t)$ for a function $g(\cdot)$ to be defined (note that $x_t$ depends only on $b_{t'}$ for $t'<t$, so this is an online algorithm). The function $g$ as well as the discount factor $\rho$ depend on the desired bound on expected loss and regret against $S_{+}$ and $S_-$. In particular, we will set $\rho=1-\frac1{T}$ for our main result on regret/loss tradeoff, and will use the freedom to choose different values of $\rho$ to obtain adaptive regret guarantees in section~\ref{sec:combine}.
The algorithm is given by

\begin{algorithm}[H]\label{alg:main}
\caption{DISCOUNTED-NORMAL-PREDICTOR}
\begin{algorithmic}[1]
\STATE $x_1\leftarrow 0$
\FOR {$t=1$ to $T$}
\STATE Predict $\text{sign}(g(x_t))$ with confidence $|g(x_t)|$. 
\STATE Set $x_{t+1}\leftarrow \rho x_t+b_t$.
\ENDFOR
\end{algorithmic}
\end{algorithm}

We start with an informal sketch of the proof, which will be made precise in Lemma~\ref{lm:potential} and Lemma~\ref{lm:g-properties} below.
The proof is based on a potential function argument. In particular, we will choose the confidence function $g(x)$ so that 
\begin{equation*}
\Phi_t=\int_{0}^{x_{t}}g(s)ds.
\end{equation*}
We will chose $g(x)$ to be an odd function, and hence will always have $\Phi_t\geq 0$.

is a potential function, which serves as  a repository for guarding our loss. 
In particular, we will choose $g(x)$ so that the change of $\Phi_t$  lower bounds the payoff of the algorithm. 
If we let $\Phi_t=G(x_t)$ (assuming for sake of clarity that $x_t>0$), where
\begin{equation*}
G(x)=\int_0^{x} g(s)ds,
\end{equation*}
we have
\begin{equation*}
\begin{split}
\Phi_{t+1}-\Phi_t&=G(x_{t+1})-G(x_t)
\approx G'(x)\Delta x+G''(x)\Delta x^2/2\approx g(x)\left[(\rho-1) x+b_t\right]+g'(x)/2.
\end{split}
\end{equation*}

\begin{figure}
\begin{center}
\begin{tikzpicture}[y=.5cm, x=0.8cm,font=\sffamily, scale=0.6]

	\draw[->] (0,0) -- coordinate (x axis mid) (10,0);
    	\draw[->] (0,0) -- coordinate (y axis mid) (0,5);
	\draw (0,0) -- coordinate (x axis mid) (-10,0);
    	\draw (0,0) -- coordinate (y axis mid) (0,-5);

	\draw[semithick] (7,5) --  (9,5);	
	\draw[semithick] (-7,-5) --  (-9,-5);	
		
	\draw[dashed] (7,-5) --  (7,5);	
	\draw[dashed] (-7,-5) --  (-7,5);

   	\draw [thick]  
   (-7,-5) to [out=100,in=-180] (0,0) 
   to [out=0,in=280] (7,5);
   
   	\draw [thick,dotted]  
   (-9,-5) to [out=0,in=215] (0,0) 
   to [out=45,in=180] (9,5);

         \draw(9, -1) node {\large $x$};

         \draw(8, +0.5) node {$+U$};
          \draw(-8, +0.5) node {$-U$};

         \draw(0.5, +4.5) node {$+1$};
          \draw(-0.5, -4.5) node {$-1$};

           \draw(-2., 1.0) node  { \large $g(x)$};
           \draw(3., 4.7) node  { \large $\tanh\left(\frac{x}{U}\right)$};

\end{tikzpicture}
\end{center}
\caption{The shape of the confidence function $g(x)$ (solid line) and the $\tanh(x)$ function (dotted line).}
\label{fig:1}
\end{figure}
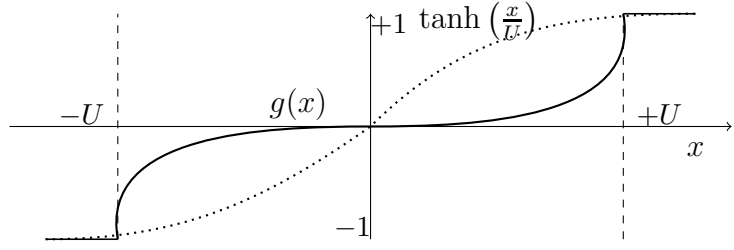

Since the payoff of the algorithm at time step $t$ is $g(x_t)b_t$, we have
\begin{equation*}
\begin{split}
\Delta\Phi_t-g(x_t)b_t=-g(x_t)(1-\rho)x_t+g'(x_t)/2,
\end{split}
\end{equation*}
so the condition becomes
\begin{equation*}
\begin{split}
-g(x_t)(1-\rho)x_t+g'(x_t)/2\leq Z,
\end{split}
\end{equation*}
where $Z$ is the desired bound on per step loss of the algorithm. Solving this equation yields a function of the form
\begin{equation*}
\begin{split}
g(x)=(2Z \sqrt{T})\cdot \erf\left(\frac{x}{4\sqrt{T}}\right)e^{x^2/(16T)},
\end{split}
\end{equation*}
where $\erf(x)=\frac{2}{\sqrt{\pi}}\int_0^{x} e^{-s^2}ds$ is the error function (see Figure~\ref{fig:1} for the shape of $g(x)$).

We now make this proof sketch precise. For $t=1,\ldots, T$ define $\Phi_t=\int_{0}^{x_{t}} g(x)dx$. The function $g(x)$ will be chosen to be a continuous odd function that is  equal to $1$ for $x>U$ and to $-1$ when $x<-U$, for some  $0<U<T$. Thus, we will have that $ |x_t|-U\leq \Phi_{t}\leq |x_t|$. Intuitively, $\Phi_t$ captures the imbalance between the number of $-1$'s and $+1$'s in the sequence up to time $t$.

 We will use the following parameters. We always have $\rho=1-1/n$ for some $n>1$ and use the notation $\bar \rho=1-\rho$. We will later choose $n=T$ to prove Theorem~\ref{thm:main}, but we will use different value of $n$ for the adaptive regret guarantees in section~\ref{sec:combine}. 
 
 We now prove that if the function $g(x)$ approximately satisfies a certain differential equation, then $\Phi_t$ defined as above is a potential function. The statement of Lemma~\ref{lm:potential} involves a function $h(x)$ that will be chosen as a step function that is $1$ when $x\in [-U, U]$ and $0$ otherwise. 

\begin{lemma} \label{lm:potential}
Suppose that the function $g(x)$ used in DISCOUNTED-NORMAL-PREDICTOR (Algorithm~\ref{alg:main}) satisfies
\begin{equation*}
\frac1{2}(\bar \rho |x|+1)^2\cdot \max_{s\in [\rho x-1, \rho x+1]} |g'(s)|\leq\bar \rho x g(x)h(x)+Z'
\end{equation*}
for a function $h(x), 0\leq h(x)\leq 1,\forall x$, for some $Z'>0$. Then
the payoff of the algorithm is at least 
\begin{equation*}
\sum_{t=1}^T \bar \rho x_t g(x_t)(1-h(x))+\Phi_{T+1}-Z'T
\end{equation*}
as long as $|b_t|\leq 1$ for all $t$.
\end{lemma}
\begin{proof}
We will show that at each $t$ 
\begin{equation*}
\Phi_{t+1}-\Phi_t\leq b_t g(x_t)+Z'-\bar \rho x_t g(x_t) (1-h(x_t)),
\end{equation*}
i.e.
\begin{equation*}
\sum_{t=1}^T b_t g(x_t)\geq -Z'T+\sum_{t=1}^T\bar \rho x_tg(x_t)(1-h(x_t))+\Phi_{T+1}-\Phi_1,
\end{equation*}
thus implying the claim of the lemma since $\Phi_1=0$.

We consider the case $x_{t}>0$. The case $x_{t}<0$ is analogous.  In the following derivation we will write $[A, B]$ to denote $[\min\{A, B\}, \max\{A, B\}]$.
\begin{description}
\item[$0\leq b_t\leq 1$:] We have $x_{t+1}=\rho x_{t}+b_t=x_t-\bar \rho x_t+b_t$, and the expected payoff of the algorithm is $g(x_t)b_t$.~Then
\begin{equation*}
\begin{split}
\Phi_{t+1}-\Phi_t&=\int_{x_t}^{x_t-\bar \rho x_t+b_t}g(s)ds\leq g(x_t)(-\bar \rho x_t+b_t)++\frac1{2}(\bar \rho x_t+b_t)^2\cdot \max_{s\in [x_t, x_t-\bar \rho x_t+b_t]} |g'(s)|\\
&\leq g(x_t)b_t+\left[-g(x_t)\bar \rho x_t++\frac1{2}(\bar \rho x_t+b_t)^2\cdot \max_{s\in [x_t, x_t-\bar \rho x_t+b_t]} |g'(s)|\right]\\
&\leq g(x_t)b_t+(-1+h(x_t))\bar \rho x_t g(x_t)+Z'.
\end{split}
\end{equation*}

\item[$-1\leq b_t\leq 0$:] This case is analogous.
\end{description}
\end{proof}

We now define $g(x)$ to satisfy the requirement of Lemma~\ref{lm:potential}.
For any \mbox{$Z, L>0$ and let }
\begin{equation}\label{eq:def-g}
g(x)=\text{sign}(x_t)\cdot \min\left\{Z \cdot \erf\left(\frac{|x|}{4L}\right) e^{\frac{x^2}{16L^2}}, 1\right\}.
\end{equation}
One can show that one has $g(x)=1$ for $|x|\geq U$ for some $U\leq 7L\sqrt{\log (1/Z)}$ (see Fact~\ref{fact-2} below). A plot of the function $g(x)$ is given in Figure~\ref{fig:1}. 

We choose 
\begin{equation} \label{eq:h-def}
h(x)=\left\{
\begin{array}{cc}
1, &|x|<U\\
0&\text{o.w.}
\end{array}\right..
\end{equation}

The following lemma shows that the function $g(x)$ satisfies all required properties stated in Lemma~\ref{lm:potential}:
\begin{lemma} \label{lm:g-properties}
Let $L>0$ be such that $\bar \rho=1/n\geq 1/L^2$. Then for $n\geq 80\log(1/Z)$ the function $g(x)$ defined by \eqref{eq:def-g} satisfies
\begin{equation*}
\frac1{2}(\bar \rho |x|+1)^2\cdot \max_{s\in [\rho x-1, \rho x+1]} |g'(s)|\leq\bar \rho x g(x)h(x)/2+2\bar  \rho L Z
\end{equation*}
for all $x$, where $h(x)$ is the step function defined above.
\end{lemma}
The intuition behind the Lemma is very simple. Note that $s\in [\rho x-1, \rho x+1]$ is not much further than $1$ away from $x$, so $g'(s)$ is very close to $g'(x)=(\frac{x}{2L^2})g(x)+\frac{1}{\sqrt{\pi} L}Z$. Since $\bar \rho\geq 1/L^2$, we have $g'(x)\leq \bar \rho x g(x)/2+\frac1{\sqrt{\pi}}\bar \rho L Z$. We now give the details of the proof.

We will need some useful properties of the function $g(x)$.
\begin{fact}\label{fact-2}
One has $g(x)=1$ for $|x|\geq 7L\sqrt{\log(1/Z)}$ as long as $Z\leq 1/e$.
\end{fact}
\begin{fact}\label{fact-1}
The function $g(x)=Z \erf \left(\frac{x}{4L}\right) e^{\left(\frac{x}{4L}\right)^2}$ is monotonically increasing and convex for any $Z>0$ for $x\geq 0$.
\end{fact}
\begin{lemma}\label{lm:g-prec}
For any $|\Delta|\leq 2$ any  $0\leq |x|\leq 2U$ one has 
\begin{equation*}
|g'(x+\Delta)|\leq 1.8 |g'(x)|
\end{equation*}
as long as $L^2\geq 80\log (1/Z)$ and $Z\leq 1/e$.
\end{lemma}
The proofs of these statements are mostly technical and are given in Appendix~\ref{app:A}.

We now give 

\begin{proofof}{Lemma~\ref{lm:g-properties}}
First, by definition of $g(x)$ we have 
\begin{equation*}
g'(x)=\bar \rho x g(x)/2+\frac{2}{\sqrt{\pi}} \bar \rho L Z ,
\end{equation*}
and together with the convexity of $g(x)$ and $g'(x+2)\leq 1.8 g'(x)$ we get 
\begin{equation*}
\frac1{2}(\bar \rho x+1)^2\cdot \max_{s\in [\rho x-1, \rho x+1]} |g'(s)|\leq |g'(x+\bar \rho x+1)| (\bar \rho x +1)^2/2 \leq \bar \rho x g(x)/2+2\bar \rho L Z.
\end{equation*}

\end{proofof}

We can now lower bound  the payoff of DISCOUNTED-NORMAL-PREDICTOR. We will use the notation
\begin{equation*}
\left|x\right|^+_{\e}=\left\{
\begin{array}{cc}
0,&|x|<\e\\
|x|-\e,&|x|>\e
\end{array}
\right.
\end{equation*}
\begin{theorem}\label{thm:smooth}
Let $n$ be the window size parameter of DISCOUNTED-NORMAL-PREDICTOR. Then one has
\begin{equation*}
\begin{split}
A_T\geq &\sum_{t=1}^{T} \bar \rho |x_t|^+_{U}+|x_{T+1}|^+_{U} -2 ZT/\sqrt{n}.
\end{split}
\end{equation*}

\end{theorem}
\begin{proof}
By Lemma~\ref{lm:g-properties} we have that the function $g(x)$ satisfies the conditions of Lemma~\ref{lm:potential}, and so from the bounds stated in Lemma~\ref{lm:potential}  the payoff of the algorithm is at least 
\begin{equation*}
\begin{split}
\sum_{t=1}^{T} \bar \rho|x_t|^+_{U}+\Phi_{T+1}-2ZT/\sqrt{n}.
\end{split}
\end{equation*}
By definition of $\Phi_t$, since $|g(x)|=1$ for $|x|\geq U$, one has $\Phi_{T+1}\geq |x_{T+1}|^+_U$, which gives the desired statement.
\end{proof}

Now, setting $n=T$, we obtain 
\begin{theorem}
\begin{equation*}
 A_T\geq \max\left\{\left|\sum_{j=1}^T b_j\right|-14\sqrt{T\log (1/Z)}, 0\right\}-2Z \sqrt{T}.
\end{equation*}
\end{theorem}
\begin{proof}
In light of Theorem~\ref{thm:smooth} it remains to bound $\sum_{t=1}^{T} \bar \rho x_t+x_{T+1}$. 
We have
\begin{equation}\label{eq:sum-phi}
\begin{split}
&\bar \rho \sum_{t=1}^{T} x_t+x_{T+1}=\bar \rho \sum_{t=1}^{T-1}\sum_{j=1}^t\rho^{t-j} b_j+x_{T+1}=\sum_{t=1}^{T-1}b_t (1-\rho^{T-t})+\sum_{t=1}^{T} \rho^{T-t} b_t =\sum_{t=1}^{T}b_t.
\end{split}
\end{equation}

Thus, since $U\leq 2\sqrt{T\log(1/Z)}$, and we chose $\rho=1-1/n=1-1/T$, we get the result by combining Theorem~\ref{thm:smooth} and equation \eqref{eq:sum-phi}.
\end{proof}

\begin{proofof}{Theorem~\ref{thm:main}}
Follows by setting $\log (1/Z)=\e^2 T$.
\end{proofof}

Note that if $Z=o(1/T)$, then the payoff of the algorithm is positive whenever the absolute value of the deviation $x_{t}$ is larger than, say $4\sqrt{n\log T}$  {\em in at least one window of size $n$}.  

We will now show that our loss/regret tradeoff is optimal up to constant factors 

\begin{proofof}{Theorem~\ref{thm:lower-bound}}
Let $A$ be an algorithm with regret at most $\sqrt{T\ln(1/Z)}$ with respect to $S_+$. Consider a sequence $X_t=\mathbf{Ber}(\pm 1, 1/2)$ of independent random variables. The payoff of $S_+$ is equal to $\sum_{t=1}^T X_t$. Since for some constant $c>0$
\begin{equation*}
\prob\left[\sum_{t=1}^T X_t>2\sqrt{T\ln (1/Z)}\right]\geq Z^c,
\end{equation*} 
we have that $A$ gets payoff at least $\sqrt{T\ln (1/Z)}$ with probability at least $Z^c$. Since the expected payoff of any algorithm on this sequence is equal to $0$, $A$ incurs loss at least $Z^c\sqrt{T\log (1/Z)}/(1-Z^c)$ on at least one sequence. This gives the statement of the theorem after choosing $Z'=Z^c$ for a suitable constant $c>0$.
\end{proofof}

Regret bounds based on different norms of the payoff sequence and non-uniform discount factors are examined in section~\ref{sec:scales+norms}.

\section{Combining strategies (lossless boosting)}\label{sec:combine}
In the previous section we derived an algorithm for the bit prediction problem with low regret to the $S_+$ and $S_-$ strategies and exponentially small loss. 
We now show how our techniques yield an algorithm that has low regret to the best of $N$ bit prediction strategies $S_1,\ldots, S_N$ and exponentially small loss.
However, since the proof works for the general experts problem, where loss corresponds to regret to a `special' expert $S_0$, we state it in the general experts setting. In what follows we will refer to regret to $S_0$ as loss. We will also prove optimal bounds on regret that hold {\em in every window of length $n$} at the end of the section. We start by proving

{\bf Theorem~\ref{thm:n-exp}}{\em 
For any $Z<1/e$ there exists an algorithm for combining $N$ strategies that has regret $O(\sqrt{T\log (N/Z)})$ against the best of $N$ strategies and loss at most $O(ZN\sqrt{T})$ with respect to any strategy $S_0$ fixed a priori. These bounds are optimal up to constant factors. 
}

We first fix notation. A {\em prediction strategy} $S$ given a bit string $b_t$, produces a sequence of weights $w_{jt}$ on the set of experts $j=1,\ldots,N$ such that $w_{jt}$ depends only on $b_{t'}, t'<t$ and $\sum_{j=1}^N w_{jt}=1, w_{jt}\geq 0$ for all $t$. Thus, using strategy $S$ amounts to using expert $j$ with probability $w_{j,t}$ at time $t$, for all $t=1,\ldots,T$. For two strategies $S_1,S_2$ we  write $\alpha_t S_1+(1-\alpha_t)S_2$ to denote the strategy whose weights are a convex combination of weights of $S_1$ and $S_2$ given by coefficients $\alpha_t\in[0, 1]$.  For a strategy $S$ we denote its payoff at time $t$ by $s_t \in [-1,+1]$. 

We start with the case of two strategies $S_1,S_2$. Our algorithm will consider $S_1$ as the base strategy (corresponding to the null strategy $S_0$ in the previous section) and will use $S_2$ to improve on $S_1$ whenever possible, without introducing significant loss over $S_1$ in the process.  We define
\begin{equation*}
\bar g(x)=\left\{
\begin{array}{cc}
g(\frac1{2}x),&x>0\\
0&\text{o.w,}
\end{array}
\right.
\end{equation*}
i.e. we are using a one-sided version of $g(x)$. 
It is easy to see that $\bar g(x)$ satisfies the conditions of Lemma~\ref{lm:potential} with $h(x)$ as defined in \eqref{eq:h-def}.
The intuition behind the algorithm is that since the difference in payoff obtained by using $S_2$ instead of $S_1$ is given by $ (s_{2, t}-s_{1,t})$, it is sufficient to 
emulate Algorithm~\ref{alg:main} on this sequence.  In particular, we set $x_t=\sum_{j=1}^{t-1}\rho^{t-1-j}(s_{2,j}-s_{1, j})$ and predict $\bar g(x_t)$ (note that since $|s_{1,t}-s_{2,t}|\leq 2$, we need to use $g(\frac1{2} x)$ in the definition of $\bar g$ to scale the payoffs).  Predicting $0$ corresponds to  using $S_1$, predicting $1$ corresponds to using $S_2$ and fractional values correspond to a convex combination of $S_1$ and $S_2$.

Formally, the algorithm $\text{COMBINE}(S_1,S_2,\rho)$ takes the following form:

\begin{algorithm}[H]\label{alg:combine}
\begin{algorithmic}[1] 
\STATE Input: strategies $S_1,S_2$
\STATE Output: strategy $S^*$
\STATE $x_1\leftarrow 0$
\FOR {$t=1$ to $T$}
\STATE Set $S^*_t\leftarrow S_{1,t}(1-\bar g(x_t))+S_{2,t}\bar g(x_t)$.
\STATE Set $x_{t+1}\leftarrow \rho x_t+ (s_{2,t}-s_{1,t})$.
\ENDFOR
\RETURN $S^*$
\end{algorithmic}
\caption{COMBINE$(S_1,S_2,\rho)$}
\end{algorithm}
Note that $\text{COMBINE}(S_1,S_2,\rho)$ is an online algorithm, since $S^*_t$ only depends on $s_{1,t'},s_{2,t'}, t'<t$.
\begin{lemma}\label{lm:combine2}
There exists an algorithm that given two strategies $S_1$ and $S_2$ gets payoff at least
\begin{equation*}
\begin{split}
\sum_{t=1}^T s_{1, t}+\max\left\{\sum_{t=1}^T (s_{2,t}-s_{1,t})-O\left(\sqrt{T\log (1/Z)}\right), 0\right\}-O(Z\sqrt{T}).
\end{split}
\end{equation*}
\end{lemma}
\begin{proof}
Use Algorithm~\ref{alg:combine} with $\rho=1-1/T$. This amounts to applying Algorithm~\ref{alg:main} to the sequence $(s_{2,t}-s_{1,t})$, so the guarantees follow by Theorem~\ref{thm:smooth}.
\end{proof}

We emphasize the property that Algorithm~\ref{alg:combine} combines two strategies $S_1$ and $S_2$, improving on the performance of $S_1$ using $S_2$ whenever possible, \emph{essentially without introducing any loss with respect to $S_1$}. Thus, this amounts to {\em lossless boosting} of one strategy's performance using another.

Algorithm~\ref{alg:combine} can be used recursively to combine $N$ strategies $S_1,\ldots, S_N$ by using a binary tree $\mathcal{T}$ with $S_j$ at its leaves. Each interior node $u\in \mathcal{T}$ can run Algorithm~\ref{alg:combine} using the left child as $S_1$ and the right child as $S_2$ as specified in Algorithm~\ref{alg:combine-all}. 

\begin{algorithm}[H]\label{alg:combine-all}
\caption{$\get(u)$}
\begin{algorithmic}[1]
\IF {$v$ is a leaf}
\RETURN $S_{v}$
\ELSE
\STATE $S_l\leftarrow \get(\text{left}(v))$
\STATE $S_r\leftarrow \get(\text{right}(v))$
\STATE $S_v\leftarrow \text{COMBINE}(S_l,S_r,\rho) $
\RETURN $S(v)$
\ENDIF
\end{algorithmic}
\end{algorithm}

\begin{theorem}\label{thm:combine-tree}
Let $S_1,\ldots, S_N$ be strategies and let $\mathcal{T}$ be a binary tree with $S_{j}, j=1,\ldots, N$ at the leaves. Denote the number of left transitions on the way from the root to $S_j$ by $d_j^l$ and the number of right transitions by $d_j^r$. Then for any $\e>4/\sqrt{T}$ there exists an algorithm that satisfies
\begin{equation*}
A_T \geq S_{j,T}- d_j^r Z\sqrt T-(d_j^r+d_j^l) \sqrt{T\log(1/Z)}
\end{equation*}
for all $j=1,\ldots, N$. This can be achieved by a convex combination of strategies $S_1,\ldots, S_N$. Note that if $S_1$ is the leftmost child, then the regret with respect to $S_1$ is exponentially small.
\end{theorem}
\begin{proofof}{Theorem~\ref{thm:combine-tree}}
Run Algorithm~\ref{alg:combine-all} on $\mathcal{T}$. The guarantees follow using Lemma~\ref{lm:combine2}. Note that the regret with respect to $S_j$ is given by the number of right transitions from the root to $S_j$ times $\sqrt{T\log(1/Z)}$, and the loss is given by $Z\sqrt{T}$ times the level of $S_j$ in $\mathcal{T}$.
 \end{proofof}

By using a specific tree structure we get

\begin{proofof}{Theorem~\ref{thm:n-exp}}
Use Algorithm~\ref{alg:combine} repeatedly to combine $N$ strategies $S_1,\ldots, S_N$ by initializing
$S^0\leftarrow S_0$  and setting $S^j \leftarrow \text{COMBINE}(S^{j-1}, S_j, 1-1/T), j=1,\ldots, N$, where $S_0$ is the null strategy. The regret and loss guarantees follow by Lemma~\ref{lm:combine2}. Thus we are using a specific (very unbalanced) tree structure of depth $N$.
\end{proofof}

\begin{corollary}\label{rmk:lognt}
Setting $Z=(NT)^{-1-\gamma}$ for $\gamma>0$,  we get  regret $O(\sqrt{\gamma T(\log N+\log T)})$ to the best of $N$ strategies and loss at most $O((NT)^{-\gamma})$ wrt strategy $S_0$ fixed a priori. These bounds are optimal and improve on the work on \cite{kearns-regret}. 
\end{corollary}

A non uniform setting of the $Z$ values in the tree structure lets us prove Theorem~\ref{thm:kol}.

\begin{proofof}{Theorem~\ref{thm:kol}}
Consider all possible Turing machines of length at most $T$ in lexicographic order (note that complexity of a bit  string of length $T$ is at most $T$.) We can view the $j^{th}$ machine whose code is simply the bits of $j$ (that is supposed to produce the bit sequence) as prediction strategies $S_j$ which predicts according to the string it outputs (if any).$comp(S_j)$, the number of bits needed to specify  $S_j$  is $O(\log j)$ since the Turing machines are lexicographically ordered; we will denote this by $\text{comp}(S_j)$. We will choose the same unbalanced tree as before with $S_j$ at its leaves so that $d_j^l=j$ and $d_j^r=1$. However will use different values for $Z$ at different depths. At depth $j$ while combining $S_j$ we will  use $Z_j=1/j^2$ at the $j$-th comparison node.  Thus the regret with respect to $S_j$ is at most $\sum_{i=1}^j \sqrt T/i^2 + \sqrt{T\log(j^2)} = O(\sqrt{T\text{comp}(S_j)}$ where we used the fact that $\sum_{i=1}^j1/i^2=O(1)$.
Thus, we get an algorithm that is simultaneously competitive against all strategies
\begin{equation*}
A_T \geq A_T(S_j)-O\left(\sqrt{T\text{comp}(S_j)}\right)
\end{equation*}
The loss can be further reduced to $1/T^{\Omega(1)}$ by combining the above tree with the null strategy $S_0$ that never bets.
\end{proofof}

\section{Regrets at multiple time scales and regret in terms of higher norms}\label{sec:scales+norms}

So far we have used $\rho=1-1/T$ for all results.  One can obtain optimal adaptive guarantees by performing boosting over a range of decay parameters $\rho$. 
 In particular, choose $\rho_j=1-1/n_j$, where $n_j, j=1,\ldots, W$ are powers of two between $80\log(NT)$ and $T$. Then let 

\begin{algorithm}[H]\label{alg:combinew}
\begin{algorithmic}[1] 
\STATE $S^{0, W}\leftarrow S_0$
\FOR {$j=W$ downto $1$}
\FOR {$k=1$ to $N$}
\STATE $S^{k,j}\leftarrow \text{COMBINE}(S^{k-1,j}, S_k, 1-1/n_j)$
\ENDFOR
\STATE $S^{0, j-1}\leftarrow S^{N, j}$
\ENDFOR
\STATE $S^*\leftarrow S^{0, 0}$
\RETURN $S^{*}$
\end{algorithmic}
\caption{Boosting over different time scales}
\end{algorithm}

We note that it is important that the outer loop in Algorithm~\ref{alg:combinew} goes from large windows down to small windows. In section~\ref{sec:k-shifting} we show another adaptive regret property of Algorithm~\ref{alg:combinew}. 

Our analysis of Algorithm~\ref{alg:combinew} requires a modification to the update rule for bounded loss prediction, and hence a slightly different version of the algorithm for combining two strategies. We now give the definitions. 

\begin{algorithm}[H]\label{alg:main-a}
\caption{DISCOUNTED-NORMAL-PREDICTOR($\rho_t$)}
\begin{algorithmic}[1]
\STATE $x_1\leftarrow 0$
\FOR {$t=1$ to $T$}
\STATE Predict $\text{sign}(g(x_t))$ with confidence $|g(x_t)|$. 
\STATE Set $x_{t+1}\leftarrow UPDATE(x_t,b_t, \rho_t)$.
\ENDFOR
\end{algorithmic}
\end{algorithm}

\begin{algorithm}[H]\label{alg:combine-a}
\begin{algorithmic}[1] 
\STATE Input: strategies $S_1,S_2$
\STATE Output: strategy $S^*$
\STATE $x_1\leftarrow 0$
\FOR {$t=1$ to $T$}
\STATE Set $S^*_t\leftarrow S_{1,t}(1-\bar g(x_t))+S_{2,t}\bar g(x_t)$.
\STATE Set $x_{t+1}\leftarrow UPDATE(x_t, s_{2,t}-s_{1,t}, \rho_t)$.
\ENDFOR
\RETURN $S^*$
\end{algorithmic}
\caption{COMBINE$(S_1,S_2,\rho_t)$}
\end{algorithm}

Here we use the function $UPDATE(x_t,b_t, \rho_t)$, which returns $\rho_t x_t+b_t$, i.e. uses discounting factors that in general depend on the time step. Note that the exact form of $\rho_t$ in Algorithm~\ref{alg:combine-a} is not specified. Different setting of $\rho_t$ discussed below yield different guarantees.

Furthermore, we will use the modified update rule given below, which only updates the deviation $x_t$ when the confidence of the algorithm is low, or when the algorithm predicts incorrectly. 
Intuitively, this ensures that the potential $\Phi_t$ never exceeds $O(\sqrt{n\log (1/Z)})$ when applied with window of size $n$, allowing us to prove regret bounds in any time window. The function $UPDATE(x_t, b_t, \rho_t)$ is given by

\begin{algorithm}[H]\label{alg:update}
\caption{UPDATE($x_t, b_t, \rho_t$)}
\begin{algorithmic}[1]
\IF {$|x_t|<U(\rho_t) \vee g(x_t)b_t<0$}
\RETURN $\rho_t x_t+b_t$
\ELSE
\RETURN $\rho_t x_t$
\ENDIF
\end{algorithmic}
\end{algorithm}
Note that the upper threshold $U$ depends on the discounting factor $\rho_t$. In what follows we will use different $\rho$ that  do not depend on time to obtain multiscale regret bounds (in this case $U(\rho)$ will be given the usual expression assuming that $\rho=1-1/n$), and also use $\rho_t=1-|b_t|^p/n$ to obtain regret bounds that depend on higher norms of the input sequence (in this case $U(\rho_t)$ will again be given by the usual expression as a function of $n$).

It is easy to see that the same regret and loss bounds hold for the algorithm that uses this update function. Indeed,
Let 
\begin{equation*}
b^*_t=\left\lbrace
\begin{array}{cc}
b_t,&\text{~if line 2 of Algorithm~\ref{alg:update} is executed at time }t\\
0&\text{o.w.}
\end{array}\right.
\end{equation*}
Let $b^0_t:=b_t-b^*_t$. Note that running DISCOUNTED-NORMAL-PREDICTOR$(\rho_t)$(Algorithm~\ref{alg:main-a}) with the new update  is equivalent to running DISCOUNTER-NORMAL-PREDICTOR(Algorithm~\ref{alg:main}) on $b^0$ with the update $x_{t+1}=\rho_t x_t+b_t$ and additionally getting all payoff from $b^*$. Thus, all loss and regret bounds of DISCOUNTED-NORMAL-PREDICTOR apply. Additionally, we now have that $|x_t|\leq U(\rho_t)+1$ for all $t$ (this will be important later). A similar argument shows that the bounds proved previously for the simple of version of the COMBINE$(S_1, S_2, \rho)$ algorithm (Algorithm~\ref{alg:combine}) hold for Algorithm~\ref{alg:combine-a}.

The analysis relies on the following two Lemmas, which are analogous to Lemma~\ref{lm:potential} and Lemma~\ref{lm:g-properties} in the paper.

\begin{lemma}\label{lm:potential-a}
Suppose that the function $g(x)$ used in Algorithm~\ref{alg:main-a} satisfies
\begin{equation*}
\frac1{2}(\bar \rho_t |x|+|b_t|)^2\cdot \max_{s\in [x, x-\bar \rho_t x+b_t]} |g'(s)|\leq\bar \rho_t x_t g(x_t)h(x_t)/2+Z'
\end{equation*}
for a function $h(x), 0\leq h(x)\leq 1$,$\forall t$, $Z'>0$ and $\bar \rho_t\geq |b_t|^2/n$.  Also, suppose that 
\begin{equation*}
\bar\eta \int_0^z g(s)ds\leq \bar \rho_t z|g(z)|/2
\end{equation*}
for all $z\leq U+1$ and all $t$. 

Then
\begin{enumerate}
\item the payoff of the algorithm satisfies
\begin{equation*}
\sum_{j=1}^t g(b_j)b_j\geq \sum_{j=1}^t \bar \rho_j x_j g(x_j)(1-h(x))+\Phi_t-Z't
\end{equation*}
\item if $\rho_t\equiv \rho$, then at each time step $1\leq t\leq T$ the $\eta$-smoothed payoff of the algorithm satisfies
\begin{equation*}
\sum_{j=1}^t\eta^{t-j} g(b_j)b_j\geq \Phi_t-Z'/(1-\rho).
\end{equation*}
\end{enumerate}
as long as $|b_t|\leq 1$ for all $t$.

\end{lemma}

\begin{proof}
We will show that at each $t$ 
\begin{equation*}
\Phi_{t+1}-\eta\Phi_t\leq b_t g(x_t)+Z'-\bar \rho x_t g(x_t) (1-h(x_t)).
\end{equation*}
Then, the result will follow from 
\begin{equation*}
\begin{split}
\sum_{j=1}^t \eta^{t-j}(\Phi_{j+1}-\rho \Phi_{j})=\Phi_{t+1}-\eta^t\Phi_1\\
\leq \sum_{j=1}^t \eta^{t-j} b_{j}g(x_{j})+Z'/(1-\rho)-\sum_{j=1}^t\rho^{t-j}\bar \rho x_{j} g(x_{j})(1-h(x_{j})),
\end{split}
\end{equation*}
and the fact that $\Phi_1=0$.

We consider the case $x_{t}>0$. The case $x_{t}<0$ is analogous.  In the following derivation we will write $[A, B]$ to denote $[\min\{A, B\}, \max\{A, B\}]$. 
\begin{description}
\item[$0\leq b_t\leq 1$:] We have $x_{t+1}=\rho x_{t}+b_t=x_t-\bar \rho x_t+b_t$, and the expected gain of the algorithm is $g(x_t)b_t$.
We have 
\begin{equation*}
\begin{split}
\Phi_{t+1}-\eta\Phi_t&=\int_{x_t}^{x_t-\bar \rho x_t+b_t}g(s)ds\\
&\leq g(x_t)(-\bar \rho x_t+b_t)+\bar \eta \Phi_t+\frac1{2}(\bar \rho |x_t|+|b_t|)^2\cdot \max_{s\in [x_t, x_t-\bar \rho_t x_t+b_t]} |g'(s)|\\
&\leq g(x_t)(-\bar \rho x_t+b_t)+\bar \rho x_{t} g(x_{t})/2+\frac1{2}(\bar \rho |x_t|+|b_t|)^2\cdot \max_{s\in [x_t, x_t-\bar \rho x_t+b_t]} |g'(s)|\\
&\leq g(x_t)b_t+\left[-g(x_t)\bar \rho x_t/2+\frac1{2}(\bar \rho |x_t|+|b_t|)^2\cdot \max_{s\in [x_t, x_t-\bar \rho x_t+b_t]} |g'(s)|\right]\\
&\leq g(x_t)b_t+(1/2)(-1+h(x_t))\bar \rho x_t g(x_t)+Z'
\end{split}
\end{equation*}
as required.

\item[$-1\leq b_t\leq 0$:] This case is analogous.
\end{description}
\end{proof}

The following lemma shows that the function $g(x)$ satisfies all required properties stated in Lemma~\ref{lm:potential}:
\begin{lemma} \label{lm:g-properties-a}
Let $L>0$ be such that $\bar \rho\geq \Delta^2/n, 1/n\geq 1/L^2$. Then for $n\geq 10000 \log(1/Z)$ the function $g(x)$ defined in \eqref{eq:def-g} satisfies
\begin{equation*}
\frac1{2}(\bar \rho |x|+\Delta)^2\cdot \max_{s\in [\rho x-\Delta, \rho x+\Delta]} |g'(s)|\leq\bar \rho x g(x)h(x)/2+Z,
\end{equation*}
for all $x$, where $h(x)$ is the step function defined above.
\end{lemma}
\begin{proof}
First, by definition of $g(x)$ we have 
\begin{equation*}
g'(x)=(1/n) x g(x)/2+\frac{2}{\sqrt{\pi}} (1/n) L Z,
\end{equation*}
and together with the convexity of $g(x)$ and $g'(x+2)\leq 1.8 g'(x)$ we get 
\begin{equation*}
\frac1{2}(\bar \rho x+\Delta)^2\cdot \max_{s\in [\rho x-1, \rho x+1]} |g'(s)|\leq |g'(x+\bar \rho x+1)| (\bar \rho x +\Delta)^2/2 \leq \bar \rho x g(x)/2+2\bar \rho L Z,
\end{equation*}
where in the last step we used the estimate

\begin{equation*}
\begin{split}
1.8\cdot(\bar \rho x +\Delta)^2\leq (\bar \rho n)((\bar \rho  (U/\rho+1)+\Delta)/\sqrt{\bar \rho n})^2\leq(\bar \rho n)((\bar \rho  U/(\rho\sqrt{\bar \rho n})+\bar \rho/\sqrt{\bar \rho n}+1)^2\\
\leq 1.8\cdot(\bar \rho n)((\sqrt{\bar \rho}  U/(\rho\sqrt{n})+\sqrt{\bar \rho/n}+1)^2\leq 1.8\cdot(\bar \rho n)((7\sqrt{\bar \rho}  \sqrt{n\log(1/Z)}/(\rho\sqrt{n})+\sqrt{\bar \rho/n}+1)^2\\
\leq 1.8\cdot(\bar \rho n)((7\sqrt{\bar \rho \log (1/Z)}/\rho+\sqrt{\bar \rho/n}+1)^2
\leq 2 \bar \rho n\\
\end{split}
\end{equation*}
when $n\geq 10000\sqrt{\log (1/Z)}$.

\end{proof}

In what follows we study two extensions of our basic framework. First, we prove bounds on the regret of our algorithm at multiple time scales (note that here we do not use the full generality of the previous two lemmas in that the discounting factor $\rho_t$ is independent of $t$). Second, we use the freedom to let the discounting factor depend on $t$ to achieve regret bounds with respect to higher norms of the input sequence.

\subsection{Multiple time scales}

We now study the role of the discounting parameter $\rho$ in Algorithm~\ref{alg:main-a} and prove that the algorithm takes advantage of any significant deviation of the sequence of payoffs from random in window of any size.

To simplify the exposition we will work with a normalized definition of $\tilde w_\rho$ in the rest of the paper. For a strategy $S$ define its {\em smoothed payoff} at time $t$ by
\begin{equation*} 
\widetilde{s}^{\rho}_t=(1-\rho) \sum_{j=1}^t \rho^{t-j} s_{j}.
\end{equation*}
Note that we have just normalized the earlier definition by a factor of $(1-\rho)$ so that it becomes like an average.
We first prove a lemma that relates a sequence smoothened with parameter $0<\rho_1\leq 1$ to the same sequence smoothened with any $\rho_2<\rho_1$. We have 
\begin{lemma}\label{lm:different-windows}
For any $\rho_1>\rho_2$ the $\rho_1$-smoothed payoff at time $t$ is a convex combination of $\rho_2$-smoothed payoffs at time $j\leq t$:

\begin{equation*}
\widetilde{s}^{\rho_1}_t=\frac{1-\rho_1}{1-\rho_2}\left[\widetilde{s}_t^{\rho_2}+\sum_{j<t}  \widetilde{s}^{\rho_2}_{t-j} \rho_1^{t-j-1} (\rho_1-\rho_2)\right].
\end{equation*}
\end{lemma}
\begin{proof}
We verify that the coefficients of $s_{t-j}$ in lhs and rhs coincide. The coefficient of $s_{t-j}$ in lhs is  $(1-\rho_1)\rho_1^j$. 
The coefficient in rhs is 
\begin{equation*}
\begin{split}
(1-\rho_1)\left[\rho_2^j+\sum_{t'=t-j}^{t-1}\rho_2^{t'-t+j} \rho_1^{t-t'-1}(\rho_1-\rho_2)\right]\\
=(1-\rho_1)\left[\rho_2^j+(\rho_1-\rho_2)\sum_{k=0}^{j-1} \rho_1^k\rho_2^{j-1-k}\right]=(1-\rho_1)\rho_1^j.
\end{split}
\end{equation*}
The coefficients in the rhs sum up to 
\begin{equation*}
\frac{1-\rho_1}{1-\rho_2}\left[1+\sum_{j\geq 1}\rho_1^j(\rho_1-\rho_2)\right]=\frac{1-\rho_1}{1-\rho_2}\left[1+\frac{\rho_1-\rho_2}{1-\rho_1}\right]=1.
\end{equation*}
\end{proof}

We  now prove

\noindent\textbf{Theorem~5}\textit{
Let $S^*$ denote the output of Algorithm~\ref{alg:combine-a}. Then the sequences $(s_{i, t}-s_{*, t})$ are $Z$-uniform for any $1\leq i\leq N$ at any scale $\rho\geq 1-1/O(\log(1/Z))$ as long as $Z=O((NT)^{-2})$.
}

\noindent\begin{proof}

We start by showing that $(s_{i, t}-s_{*, t})$ are $Z$-uniform at all scales $n=2^j, 1\leq j\leq \log T$, $n\geq 10000\log(1/Z)$, corresponding to discount factor $\rho=1-1/n$. Consider the application of $S_{i}$ at windows size $n=2^j$. Denote the payoff of the base strategy for $S_{i}$ at this window by $s_{0,t}$ and denote the coefficient in the convex combination by $g_t$, so that $s^{i, j}_{t}=s_{0,t}+g_t(s_{i, t}-s_{0, t})$. Then one has by Lemma~\ref{lm:g-properties-a}
\begin{equation}\label{eq:star}
\begin{split}
\sum_{j=0}^t \rho^{t-j} (s_{0, t}+g_t(s_{i, t}-s_{0, t})-s_{i, t})=\sum_{j=0}^t \rho^{t-j} (s_{i, t}-s_{0, t})(1-g_t)\\
\leq x_{t+1}-\Phi_{t+1}+Zt/\sqrt{n}\leq O(\sqrt{n\log(1/Z)})+O(Zt/\sqrt{n}),
\end{split}
\end{equation}
where we used the fact that $\sum_{j=0}^t \rho^{t-j} (s_{i, t}-s_{0, t})$ is exactly the discounted deviation $x_{t+1}$, and 
$\sum_{j=0}^t \rho^{t-j} (s_{i, t}-s_{0, t})g_t\geq \Phi_{t+1}-O(Zt/\sqrt{n})$ by Lemma~\ref{lm:g-properties-a}.

We have shown that the sequence $(s^{i,j}_t-s_{i,t})$ is $Z$-uniform at scale $2^j$ after the application of $s_{i,t}$ at level $j$, and it remains to show that this property is not destroyed by the subsequent combinations. By Lemma~\ref{lm:g-properties-a} one has that for any $t$
\begin{equation}\label{eq:star-ijt}
\sum_{j=0}^t \rho^{t-j}(s^{i,j}_t-s_{*,t})\leq -O(ZNT\log T),
\end{equation}
and hence by combining \eqref{eq:star} and \eqref{eq:star-ijt} we get
\begin{equation*}
\sum_{j=0}^t \rho^{t-j}(s_{i, t}-s_{*,t})\leq O(\sqrt{n\log(1/Z)})-O(ZNT\log T).
\end{equation*}

We now show that the sequence $(s_{i,t}-s_{*, t})$ is $Z$-uniform at any scale.
Consider a value of $\rho\neq 1-1/2^j$. Let $l>0$ be such that $\bar \rho_l/2\leq \bar \rho\leq \bar \rho_l$. Set $n_l=(1-\rho_l)^{-1}$.  By Lemma~\ref{lm:different-windows} one has for any sequence $b$
\begin{equation*}
\begin{split}
\widetilde{b}^\rho \leq \frac{1-\rho}{1-\rho^l}\left[\widetilde{b}_t^{\rho_l}+\sum_{j<t}  \widetilde{b}^{\rho_l}_{t-j} \rho_1^{t-j-1} (\rho-\rho_l)\right],
\end{split}
\end{equation*}
where the coefficients in the rhs are non-negative and sum up to $1$. Thus, setting $b=s_{i,t}-s_{*, t}$, we get the desired conclusion for all $\rho\geq 1-1/(10000\log(1/Z))$.
Thus, the discounted deviation is $O(\sqrt{n\log(1/Z)})$ as long as $Z=O((NT)^{-2})$.
\end{proof}

\begin{corollary}
Suppose that we are given a set of strategies $S_1,\ldots, S_N$ for the bit prediction problem. Then by alternating $S_{i}$ with the no prediction strategy in Algorithm~3 we can ensure that the final sequence of payoffs is essentially nonnegative in every window and is $Z$-uniform wrt each $S_i$ at any scale $\rho\geq 1-1/(80\log(1/Z))$ as long as $Z=O((NT)^{-2})$.
\end{corollary}

\begin{lemma}
One has 
\begin{equation*}
\bar\eta \int_0^z g(s)ds\leq \bar \rho zg(z)/2
\end{equation*}
for all $z\leq U+1$, as long as $\bar \eta\leq \bar \rho$, $L\geq 80\log(1/Z), Z\leq 1/e$.
\end{lemma}
\begin{proof}
First note that for $z\in [-U, U]$
\begin{equation*}
\int_0^z g(x)dx \leq z g(z)/2
\end{equation*}
due to the convexity of $g(x)$ for $x\in [-U, U]$.

Now suppose that $z\in [U, U+1]$. Note that 
\begin{equation*}
\int_0^U g(x)dx 
\end{equation*}
is maximized when $Z$ and $L$ are the smallest possible due to the convexity of $g(x)$ for $x\in [0, U]$. Since $U\geq 80$, we have 
\begin{equation*}
\frac1{U}\int_0^U g(x)dx \leq 0.3
\end{equation*}

Since $U\geq 80$, we have for $|z|\geq U$
\begin{equation*}
\int_0^{z} g(x)dx\leq \int_0^{U} g(x)dx+(|z|-|U|)\leq 0.3 |U|+|z|-|U|\leq 0.4 |z|.
\end{equation*}
This completes the proof of the lemma.
\end{proof}

\subsection{Higher norms}
In this section we prove regret bounds that depend on higher norms of the input sequence. Our main tool here is an extension of our analysis to the setting where the discount factor $\rho$ is allowed to change over time.

Note that when the discounting parameter is allowed to depend on time, the discounted deviation takes form
\begin{equation*}
x_t=\sum_{j=1}^{t-1} b_j \prod_{i=j}^{t-2}\rho_i.
\end{equation*}

The following crucial property is a more general version of \eqref{eq:sum-phi}:
\begin{lemma}\label{lm:sum-phi-disc}
\begin{equation*}
\sum_{j=1}^{T} (1-\rho_j) x_j+x_{T+1}=\sum_{j=1}^T b_j.
\end{equation*}
\end{lemma}
\begin{proof}
Induction on $T$.
\begin{description}
\item[Base case:$T=1$] The statement is true since $x_1=b_1$.
\item[Inductive step:] 
\begin{equation*}
\begin{split}
\sum_{j=1}^{T-1} (1-\rho_j) x_j+x_T
=\sum_{j=1}^{T-2} (1-\rho_j) x_j+(1-\rho_{T-1})x_{T-1}+\rho_{T-1} x_{T-1}+b_T\\
=\sum_{j=1}^{T-2} (1-\rho_j) x_j+x_{T-1}+b_T
=\sum_{j=1}^{T-1} b_j+b_T,
\end{split}
\end{equation*}
where we used the inductive hypothesis in the last step.
\end{description}
\end{proof}

Regret bounds obtained so far depend on the best upper bound on $|b_t|$ that is available: in fact, we assumed that the input is scaled so that $|b_t|\leq 1$. Thus, the bounds on the regret  scale linearly with $||b||_\infty$. This is tight up to constant factors, as shown in Theorem~\ref{thm:lower-bound}. It is natural to ask if better bounds can be obtained if the sequence $b_t$ has small $l_p$ norm for some $p>0$.  We will use the notation $\mu_p(b):=\sum_{t=1}^T b_t^p$. By choosing $\rho_t=1-|b_t|^p/n$, we get

\begin{theorem}\label{thm:p-norm-bounds}
Let $b_t$ be the sequence of payoffs such that $|b_t|\leq M$ and $\mu_p(b)\geq O(\log(1/Z))$ (this can be achieved by rescaling the values of $b_t$ if necessary, thus increasing the value of $M$).  Then one can obtain regret at most $M\sqrt{\mu_p(b) \log (1/Z)}$ and loss at most $MZ\sqrt{\mu_p(b)}$ for any $p\leq 2$.
\end{theorem}
\begin{proof}
Fix $0\leq p\leq 2$. Note that all regret and loss bounds obtained so far scale linearly with $M$. 
The loss property follows immediately from Lemma~\ref{lm:potential-a} by setting $\rho(b)=1-|b|^p/n, n=\mu^*$.

Using Lemma~\ref{lm:potential-a} and Lemma~\ref{lm:sum-phi-disc}, we get that the regret is at most 
\begin{equation*}
O\left(M\sqrt{n \log(1/Z)}+M\sqrt{n \log(1/Z)} \sum_{t}\bar \rho_t\right)=O\left(M\mu_p(b) \sqrt{n^{-1} \log(1/Z)}+M\sqrt{n\log(1/Z)}\right).
\end{equation*}
Setting $n=\mu^*$, we get regret $O\left(M\sqrt{\mu^*\log(1/Z)}\right)$.
\end{proof}
\section{Regret to $k$-shifting optima}\label{sec:k-shifting}
In this section we prove regret bounds to $k$-shifting optima, i.e. when the $T$ time steps are partitioned into intervals and a different strategy may be used in each partition. In such a case our regret is at most $\sum_j O\left(\sqrt{|I_j| \log (1/Z)}\right)$, where $|I_j|$ is the length of the $j$-th interval. In addition we also preserve the low loss property. In particular, we now turn to proving

{{\bf Theorem \ref{thm:windows-adaptive}}\em
Consider the result of running Algorithm 4 on a set of strategies $\{S_1,\ldots, S_N\}$ with a base strategy $S_0$. Let $I_1,\ldots, I_k, I_j=[A_j, B_j]\subset[1..T]$ be a covering of $[1..T]$ by disjoint intervals. Then for any assignment of strategies to intervals $\eta_j, 1\leq j\leq k$ one has
\begin{equation*}
\begin{split}
A_T&\geq\max\left\{\sum_{j=1}^k\left[S_{\eta_j}(I_j)-O\left(\sqrt{|I_{j}| \log (1/Z)}\right)\right],\text A_T(S_0)\right\}-O(ZNT\log T),
\end{split}
\end{equation*}
where $S_{\eta_j}(I)$ is the cumulative payoff of strategy $S_{\eta_j}$ on interval $I$.
}

Before we give the proof of Theorem~\ref{thm:windows-adaptive}, we prove the following 
\begin{lemma}\label{lm:combine2-a}
For all $1\leq t\leq T$ let $A_t$ denote the payoff of Algorithm~\ref{alg:combine}. Then
\begin{equation*}
\begin{split}
A_t&\geq \sum_{j=1}^t s_{1,t}+\sum_{j=1}^{t-1}{|\tilde s_{2, j}^\rho-\tilde s_{1, j}^\rho|}^{+}_{U}+{|\tilde s_{2, t}^\rho-\tilde s_{1, t}^\rho|}^{+}_{U}/\bar \rho-ZT(1-\rho)^{1/2}.
\end{split}
\end{equation*}
\end{lemma}
\begin{proof}
Algorithm~\ref{alg:combine} amounts to applying Algorithm~\ref{alg:main-a} to the sequence $(s_{2}-s_{1})$, and hence by Theorem~\ref{thm:smooth} the payoff of Algorithm~\ref{alg:combine} is at least
\begin{equation*}
\begin{split}
\sum_{j=1}^t (s_{1,j}	+(s_{2,j}-s_{1,j}) \bar g(x_j))=\sum_{j=1}^t s_{1,j}+\sum_{j=1}^t (s_{2,j}-s_{1,j}) \bar g(x_j)\\
\geq \sum_{j=1}^t s_{1,j}+\sum_{j=1}^{t-1} {| s_{2, j}^\rho-\tilde s_{1, j}^\rho|}^{+}_{U}+{|\tilde s_{2, t}^\rho-\tilde s_{1, t}^\rho|}^{+}_{U}/(1-\rho)-ZT(1-\rho)^{1/2}\\
\end{split}
\end{equation*}

This immediately yields (2), and we get (1) by setting parameters as stated.
\end{proof}

We can now give

\begin{proofof}{Theorem~\ref{thm:windows-adaptive}}

Consider the application of strategy $S_{\eta_j}$ with window size $2^{r_j}$,  where $2^{r_j}\leq |I_j|\log (1/Z)\leq 2^{r_j+1}$. 
Denote the strategy that $S_{\eta_j}$ is applied to by $S^0_{\eta_j}$. Let $\Delta_{\eta_j}=s_{\eta_j}-s^0_{\eta_j}$ be the difference of payoffs.
Then one has by Lemma~\ref{lm:combine2-a}
\begin{equation*}
\begin{split}
A_T&\geq \sum_{j=1}^k \left[\sum_{t=A_j}^{B_j}s^0_{\eta_j}+\tilde \Delta_{\eta_j, t}-{U_j}-O(ZNT\log T)\right],
\end{split}
\end{equation*}
where $U_j=O(\sqrt{2^{r_j}\ln(1/Z)})$.

We need to relate the summation of smoothed payoff differences $\tilde \Delta_{\eta_j, t}$ to the cumulative difference of payoffs $S_{\eta_j}$ and $S^0_{\eta_j}$ in $[A_j, B_j]$. We relate these quantities as follows. Fix $j$.
\begin{equation} \label{eq:sumAB}
\begin{split}
\sum_{t=A_j}^{B_j} \tilde \Delta_{\eta_j, t}=(1-\rho)\sum_{t=A_j}^{B_j}\sum_{t'=A_j}^t \rho^{t-t'} \Delta_{\eta_j, t'}+(1-\rho)\sum_{t=A_j}^{B_j}\rho^{t-A_j+1} \tilde \Delta_{\eta_j, A_j-1}\\
\geq (1-\rho)\sum_{t'=A_j}^{B_j} \Delta_{\eta_j, t'}\sum_{l=0}^{B_j-t'}\rho^{l}-|\tilde \Delta_{\eta_j, A_j-1}|=\sum_{t'=A_j}^{B_j} \Delta_{\eta_j, t'}(1-\rho^{B_j-t'+1})-|\tilde \Delta_{\eta_j, A_j-1}|\\
\geq \sum_{t'=A_j}^{B_j} \Delta_{\eta_j, t'}-|\tilde \Delta_{\eta_j, B_j+1}|-|\tilde \Delta_{\eta_j, A_j-1}|\geq \sum_{t'=A_j}^{B_j} \Delta_{\eta_j, t'}-O(U)\\
\end{split}
\end{equation}
since by assumption $|\tilde \Delta_{\eta_j, t}|=O(U_j)$ for all $t$ (guaranteed by the update rule in Algorithm~\ref{alg:update}).

Thus, we have that
\begin{equation*}
\begin{split}
A_T&\geq \sum_{j=1}^k\left[\sum_{t=A_j}^{B_j} s_{\eta_j}(t)-O(\sqrt{2^{r_j}\log (1/Z)})-O(|I_j|\sqrt{2^{-r_j}\ln(1/Z)})-O(ZNT\log T)\right]\\
&\geq \sum_{j=1}^k\left[S_{\eta_j}(I_j)-O\left(\sqrt{|I_j| \log (1/Z)}\right)\right]-O(ZNT\log T)
\end{split}
\end{equation*}
by the choice of $r_j$.
Also, one sees that $A_T\geq A_T(S_0)-O(ZNT\log T)$, i.e. we maintain the essentially zero loss property with respect to $S_0$.
\end{proofof}

It is interesting to note that our algorithm has optimal regret bounds with respect to {\em any partition} of $[1..T]$ into disjoint intervals and is completely parameter-free. Also, setting $Z=o((NT)^2)$, we obtain vanishingly small loss with respect to $S_0$, maintaining an optimal loss/regret tradeoff.

\section{Non discounted strategy cannot achieve small loss}
We will show that if we set $\rho = 1$ (that is, use  $x_t=\sum_{j=1}^{t-1}  b_j)$ , then there is no function $g$ such that predicting $g(x_t)$ with a reasonable regret and a tiny loss.
\begin{claim}\label{nodiscount}
If $\rho =1$, then for all functions $g$,  the maximum loss $L$ and regret $R$ are such that $RL \ge O(T)$.
\end{claim}
\begin{proof}
We will look the plot of $g$ on the range $x \ge 0$. Assume $g(0) \le 0$ as otherwise our arguments will apply on the negative side of the $x$-axis.
The main idea is  that  for large $x$ it must be close to $1$ to avoid a high regret. Precisely, look at the earliest point $\Delta$ such that  $g(\Delta) \ge 0.5$. Clearly $R \ge 0.5 \Delta$ for the string where the initial $\Delta$ values are $+1$ and the remaining $0$.  Now to consider the string $+1$ repeated $\Delta$ times followed by $-1$ repeated $\Delta$ times, and this pair repeated $\lfloor T/(2\Delta) \rfloor$ times (padded with $0$'s to make it of size $T$). In each such pair the algorithm loses at least $g(\Delta)$ and so $L \ge  0.5\lfloor T/(2\Delta) \rfloor$. The claim follows.
\end{proof}

\section{Risk-free assets, transaction costs and high probability bounds}\label{app:risk-free}

The results on predicting a bit sequence without loss given above yield the following construction of risk-free assets. Let $v_t$ be the expected price of a stock at time $t$ and let $x_t=\log(v_t)-\log(v_1)$. Let the expected rate of return of the stock be $r$, i.e.  $\expect[x_T]=r T$. We make an important simplifying assumption that the percentage change in the price of the stock is bounded, for example $|x_t|\leq 1$. Running our algorithm on the sequence $x_t$ produces a sequence of confidence values $g_t$, $t=1,\ldots, T$, which are interpreted as a signal to buy if $g_t>0$ and sell otherwise, where $|g_t|$ specifies the amount of stock to buy/sell. The bounded loss property now implies that this investment strategy does not lose more than $Z\sqrt{T}$ of the initial capital at any time $t=1,\ldots, T$. The regret property means that the rate of return of the investment strategy is at least $r-4\sqrt{\log (1/Z)/T}-Z/\sqrt{T}$. 

This construction assumes zero transaction cost and the ability to trade fractional amounts of shares. However, transaction costs may make these assumptions unrealistic. This motivates introducing randomness into the process and interpreting $|g_t|$ as the probability of buying/selling rather than the actual amount. The guarantees on expected regret carry over immediately, but it becomes desirable to have a high probability bound on the regret and loss in this setting.  This motivates introducing randomness into the process and interpreting $|g_t|$ as the probability of buying/selling or doing nothing rather than the actual amount that the strategy buys/sells at each point in time. The guarantees on expected regret carry over immediately, but it becomes desirable to have a high probability bound on the regret and loss in this setting. We show that our algorithm has bounded loss and good regret with high probability.

Consider the function $g(x)$ defined as follows: 
\begin{equation}\label{eq:lin-g}
g(x)=\left\{
\begin{array}{cc}
Z(x/\e T),&x<\e T\\
Ze^{(x-\e T)^2/(16T)},&\e T\leq x\leq \e T+4\sqrt{T\log(1/Z)}\\
1&\text{o.w.}
\end{array}
\right.
\end{equation}
and 
\begin{equation*}
h(x)=\left\{
\begin{array}{cc}
1,&x<\e T\\
1/2,&\e T<x<T+4\sqrt{T\log (1/Z)}\\
0&\text{o.w.}
\end{array}
\right.
\end{equation*}

It is easy to see that 
\begin{equation*}
\frac1{2}(\bar \rho |x|+1)^2\cdot \max_{s'\in [\rho x-1, \rho x+1]} g'(s')\leq\bar \rho x g(x)h(x)+Z.
\end{equation*}

We get regret at most $2\e T$ by the same arguments as in Theorem~\ref{thm:main}. We now prove the high probability bound on the loss given in Theorem~\ref{thm:high-prob}.

We prove
\begin{theorem}\label{thm:transactions}
Let $0\leq c\leq 1$ be a transaction cost. There exists a randomized strategy that yields exponentially small loss and expected regret at most $(1-4c)T$ in the presence of transaction cost of $c$ per trade.
\end{theorem}
\begin{proof}
Set $\e=2c$ in \eqref{eq:lin-g}. Note that the expected transaction costs incurred are $\sum_{t=1}^T cg(x_t)$. We have
\begin{equation*}
\begin{split}
\text{gain}&\geq \sum_{t=1}^T \bar \rho x_t g(x_t)(1-h(x_t))-\sum_{t=1}^T cg(x_t) -2c Z T\\
&\geq \sum_{t:2cT\leq x_t\leq 3cT} \left[\bar \rho x_t g(x_t)/2-cg(x_t)\right]+\sum_{t:x_t\geq 3cT}\left[\bar \rho x_t g(x_t)-cg(x_t)\right] -3c Z T\\
&\geq \sum_{t:x_t\geq 3cT}\left[\bar \rho x_t g(x_t)-cg(x_t)\right] -3c Z T\\
\end{split}
\end{equation*}
 Thus, the gain is at least $-3cZT$ even after discounting transaction costs, and the regret is at most $4cT$ by the same argument as in Theorem~\ref{thm:main}, which we do not repeat here.
\end{proof}

We also show that we can get essentially zero loss in expectation in the presence of transaction costs:
\begin{theorem}\label{thm:high-prob}
The loss of Algorithm~\ref{alg:main-a} achieving regret  $\e T$ is $O(\log (1/\delta)/\e)$ for any $1\leq t \leq T$ with probability at least $1-\delta$ for any $\delta>0$.
In particular, by setting $\delta=1/\e$ and terminating the algorithm if the loss is larger than $O(\log (1/\e)/\e)$, we get an algorithm with regret at most $4\e T$ and loss $O(\log(1/\e)/\e)$. 
\end{theorem}
\begin{proof}
Let $x_t$ be the sequence of discounted deviations, and let $X_{j}$, $1\leq j\leq T$ be the $\pm 1$ random variables corresponding to the bets that the algorithm makes. We need to show that 
\begin{equation*}
\prob\left[\sum_{j=1}^t X_{j}<-2\log (1/\delta)/\e\right]<\delta.
\end{equation*}
Define
\begin{equation*}
\mu_t:=\sum_{j=1}^t \expect[X_j]\geq \sum_{j=1}^t \bar \rho x_j g(x_j).
\end{equation*}
We also have 
\begin{equation*}
\sigma_t^2=\expect\left[\sum_{j=1}^t X_{j}^2\right]=\sum_{j=1}^t g(x_{j}).
\end{equation*}
Since the probability of making a bet when $\bar \rho x_j\leq \e$ is at most $Z<1/e$, we have that the payoff cannot be smaller than $-\log(1/\delta)$ with probability larger than $1-\delta$. Otherwise, when $\bar \rho x\geq \e$, we have $\sigma_t^2\leq \mu_t/\e$ .

We have by Bernstein's inequality
\begin{equation*}
\prob\left[\sum_{j=1}^t X_{j}<-\log(1/\delta)/\e\right]<\exp\left[\frac{(\mu_t+\log(1/\delta)/\e)^2}{\sigma^2_t+\log(1/\delta)/3\e}\right].
\end{equation*}
Since
\begin{equation*}
\frac{(\mu_t+\log(1/\delta)/\e)^2}{\sigma_t^2+\log(1/\delta)/3\e}\geq \frac{(\mu_t+\log(1/\delta)/\e)^2}{\mu_t/\e+\log(1/\delta)/3\e}\geq \log (1/\delta),
\end{equation*}
we get the desired result. The last inequality can be verified by considering two cases: $\mu>\log(1/\delta)/3$ and $\mu<\log(1/\delta)/3$.
\end{proof}

Theorem~\ref{thm:high-prob} is optimal up to constant factors:
\begin{theorem}\label{thm:high-prob-lb}
Any algorithm achieving regret  $\e T$ with $\pm 1$ betting amounts incurs loss $\Omega(\log (1/\delta)/\e)$ with probability at least $1-\delta$, for any $\delta>0$.
\end{theorem}
\begin{proof}
Let $b_t$ be iid $\textbf{Ber}(\pm 1, \frac{1+\e}2)$ and denote the confidence of the algorithm at time $t$ by $g_t$.  Let $T$ be the (random) maximum time such that 
$\sum_{j=1}^T g_j<1/\e^2$. Thus, $\sum_{j=1}^T g_j\geq 1/\e^2-1$.

One has 
\begin{equation*}
\expect_b\expect_{alg}\left[\sum_{j=1}^T g_t b_t\right]\leq 1/\e,
\end{equation*}
where $\expect_b$ denotes expectation with respect to $b$ and $\expect_{alg}$ denotes expectation wrt the randomness of the algorithm. Thus, there exists an input sequence $b^*$ for which $\expect_{alg}\left[\sum_{j=1}^T g_t b^*_t\right]\leq 1/\e$. In particular, with probability at least $1/2$ one has $\sum_{j=1}^T g_t b^*_t\leq 2/\e$.

Thus, we have that on the sequence $b^*$ the expected return of the algorithm is at most $2/\e$ with probability at least $1/2$ (over the coin flips of the algorithm).
The payoff of the algorithm is then a sum of Bernoulli variables with  expectation at most $2/\e$ and  variance  $\sum_{j=1}^T g_t\geq (1/\e^2-1)$. Thus, at time $T$ the loss  is as large as $\Omega(\log(1/\delta)/\e)$ with probability at least $\delta/2$ for any $\delta>0$.
\end{proof}

\section{Applications}\label{app:app}
In this section we show applications of our framework to two problems in online learning: the adversarial multi-armed bandit problem in the partial information model
and online optimization. 

\subsection{Partial information model}
We show that a simple application of our framework can be used to obtain an algorithm with sublinear regret and essentially zero expected loss with respect to the average of all arms. In particular, for any $Z<T^{-2}$ we will obtain an algorithm with regret $O(N^{1/3}T^{2/3}(\log(1/Z))^{1/3})$ and expected loss at most $ZT$ with respect to the average of all arms. 

Let the rewards of $N$ arms be given by $x_i(t)$, $1\leq i\leq N, 1\leq t\leq T$. We will define probabilities $p_i(t)$ of sampling arms $i$ at time $t$ inductively. Let $\gamma>0$ be a parameter to be fixed later. Let $I_t$ be the (random) arm played at time $t$. Define 
\begin{equation*}
\hat x_i(t)=\left\lbrace
\begin{array}{cc}
x_i(t)/p_i(t),&\text{~if~}I_t=i\\
0&\text{o.w.}
\end{array}
\right.
\end{equation*}

Define $p_i(t)$ as follows:
\begin{description} 
\item[$t=1$] $p_i(t)=1/N$ for all $i=1,\ldots, N$.
\item[$t\to t+1$] Consider the sequence of payoffs $\hat x_i(t)$ as a full information problem (thus, at each time step $t$ all $\hat x_i(t)$ except for possibly the one that was played are zero). Note that $|\hat x_i(t)|\leq N/\gamma$ since $p_i(t)\geq \gamma/N,\forall i,t$. Run Algorithm~\ref{alg:combine-all} on this sequence after scaling it down by a factor of $N/\gamma$ using $\rho(b_t)=1-|b_t|/n$. Let $r_i(t+1)$ be the probability of playing arm $i$ at time $t+1$ given by Algorithm~\ref{alg:combine-all}. Let $p_i(t+1)=(1-\gamma)r_i(t+1)+\gamma/N$.
\end{description}

We have 
\begin{lemma}
For any $Z<(NT)^{-2}$ the expected regret of the algorithm is at most $O(N^{1/3}T^{2/3}(\log(1/Z))^{1/3})$ and the expected loss is $O(ZNT)$.
\end{lemma}
\begin{proof}
First consider the auxiliary full information problem. By theorem~\ref{thm:p-norm-bounds} with $p=1$ the regret is at most 
\begin{equation*} 
O(U+U\sum_{t=1}^T \bar\rho_t)=O(U+U\sum_{t=1}^T (\gamma/N)\hat x_i(t)/n),
\end{equation*}

Set $n=T\gamma/N, U=\sqrt{T(\gamma/N)\log(1/Z) }$. Then the expected regret is
\begin{equation*}
\expect\left[O\left(U+U\sum_{t=1}^T (\gamma/N)\hat x_i(t)/n\right)\right]=O\left(\sqrt{T(\gamma/N)\log(1/Z) }\right),
\end{equation*}
where we used the fact that $\expect[\hat x_i(t)]=x_i(t)$.

Thus, the final expected regret is at most
\begin{equation*}
O\left(\gamma T+(N/\gamma)\sqrt{T(\gamma/N)\log(1/Z) }\right)=O\left(\gamma T+\sqrt{NT\log(1/Z)/\gamma}\right),
\end{equation*}
where $\gamma T$ comes from the fact that the algorithm pulls a uniformly random arm with probability $\gamma$.
Setting $\gamma^{3/2}=\sqrt{N\log(1/Z)/T}$, we get
\begin{equation*}
O\left(N^{1/3}T^{2/3}(\log(1/Z))^{1/3}\right).
\end{equation*}
The expected loss with respect to the average of all arms is 
\begin{equation*}
O((N/\gamma)(ZT/U))=O(ZNT)
\end{equation*}
\end{proof}

It is interesting to note that unlike the full information model, one cannot achieve essentially zero loss with respect to an arbitrary strategy in the partial information model.

\subsection{Online optimization}
We first note that our techniques yield algorithms in the online decision making framework of \cite{kalai-vempala-jcss} that have optimal regret with respect to dynamic strategies.  We do not state the guarantees here since the exposition in \cite{kalai-vempala-jcss} is quite similar to the experts problem.  One interesting consequence of our analysis that should be noted is as follows. Let $x_t, t=1,\ldots, T$ be an adversarial real-valued sequence, $x_t\in [-1,1]$ presented to the algorithm in an online fashion. Then a straightforward application of Theorem~\ref{thm:Z-rand} implies that one can approximate the signal $x_t$ by $\hat x_t$ so that the cumulative deviation of $x_t$ from $\hat x_t$ in any window of size $n=\Omega(\log T)$ is not greater than $O(\sqrt{n\log T})$, i.e. the deviation that one would expect  to see with probability $1-T^{\Theta(1)}$ if the difference were uniformly random.

We now show how our framework can be applied to online convex optimization methods of \cite{zinkevich}. We start by defining the problem.
Suppose that the algorithm is presented with a sequence of convex functions $c_t: F\subset \mathbb{R}^n\to \mathbb{R}$, $t=1,\ldots, T$. Denote the decision of the algorithm at time $t$ by $x_t$. The objective is to minimize regret against the best single decision in hindsight:
\begin{equation*}
\sum_{t=1}^T c_t(x_t)-\max_{x\in F} \sum_{t=1}^T c_t(x)
\end{equation*}

If the functions $c_t$ are convex, gradient descent methods can be used in the online setting \cite{zinkevich} to get efficient algorithms. We state the greedy projection algorithm here for convenience of the reader:

\begin{algorithm}[H]\label{alg:zinkevich}
\caption{Greedy projection algorithm (\cite{zinkevich})}
\begin{algorithmic}[1]
\STATE Select $x_1\in F$ arbitrarily, choose a sequence of learning rates $\eta_t,t=1,\ldots, T$
\FOR {$t=1$ to $T$}
\STATE Set $x_{t+1}\leftarrow P(x^t-\eta_t \nabla c^t(x_t))$.
\ENDFOR
\end{algorithmic}
\end{algorithm}
Here $P$ is the orthogonal projection operator onto $F$. In what follows we use $||F||$ to denote (an upper bound on) the diameter of $F$, and $||\nabla c||$ to denote an upper bound on the norm of the gradient of $c_t$ on $F$.

One has
\begin{theorem}(\cite{zinkevich})
The greedy projection algorithm with $\eta_t=t^{-1/2}$ has regret at most $||F||^2\sqrt{T}/2+(\sqrt{T}-1/2)||\nabla c||^2$.
\end{theorem}

The following notion introduced in \cite{zinkevich} parameterizes dynamic strategies in the online gradient descent setting:
\begin{definition}(\cite{zinkevich})
The \textbf{path length} of a sequence $x_1,\ldots, x_T$ is 
\begin{equation*}
\sum_{t=1}^{T-1} d(x^t, x^{t+1}). 
\end{equation*}
Define $\mathbb{A}(T, L)$ to be the set of sequences with $T$ vectors and path length less than $L$.
\end{definition}

\begin{definition}(\cite{zinkevich})
Given an algorithm $A$ and a maximum path length $L$, the \textbf{dynamic regret} $R_A(T, L)$ is 
\begin{equation*}
R_A(T, L)=C_A(T)-\sum_{A'\in \mathbb{A}(T, L)} C_{A'}(T).
\end{equation*}
\end{definition}

Zinkevich(\cite{zinkevich}) shows that 
\begin{theorem}(\cite{zinkevich})
If $\eta$ is fixed, the dynamic regret of the greedy projection algorithm is 
\begin{equation*}
R_G(T, L)\leq \frac{7||F||^2}{\eta}+\frac{L||F||}{\eta}+\frac{T\eta ||\nabla c||^2}{2}.
\end{equation*}
\end{theorem}

Black-box application of techniques of \cite{zinkevich} requires setting the learning rate $\eta$ to the value given by path length that one would like to be competitive against.
It would be desirable to devise an algorithm that is simultaneously competitive against all possible path lengths.
Choose $\eta_j=2^{-j}, j=1,\ldots, \log T$, $\rho_i=1-2^{-i}, i=1,\ldots, \log T$. Let $S_{i,j}$ be the strategy that applies the gradient descent algorithm with $\eta=\eta_j$, $\rho=\rho_i$.  We then have
\begin{theorem}
Choose any $Z<1/e$ and any partition of $[1:T]$ into disjoint intervals $I_j, j=1,\ldots, k$. Let the desired path length for $I_j$ be $\gamma_j ||F|| |I_j|$.
Then the regret of the tree-based comparison algorithm is at most 
\begin{equation*}
\begin{split}
\sum_{j=1}^k\left[O(||F||||\nabla c|| \gamma_j^{-1/2}|I_j|)+O\left(\sqrt{|I_{j}| \log (1/Z)}\right)\right]+O(ZT(\log T)^2).
\end{split}
\end{equation*}
\end{theorem}
\begin{proof}
Follows by Theorem~\ref{thm:windows-adaptive}.
\end{proof}

\newpage

\begin{thebibliography}{10}

\bibitem{log-online-convex}
A.~Agarwal, E.~Hazan, and S.~Kale.
\newblock Logarithmic regret algorithms for online convex optimization.
\newblock {\em Machine Learning}, 69, 2007.

\bibitem{agarwal}
R.~Agarwal.
\newblock Sample mean based index policies with $o(\log n)$ regret for the
  multi-armed bandit problem.
\newblock {\em Advances in Applied Probability}, 27:1054--1078, 1995.

\bibitem{audibert-bubeck}
J.-Y. Audibert and S.~Bubeck.
\newblock Minimax policies for adversarial and stochastic bandits.
\newblock {\em COLT}, 2009.

\bibitem{auer}
P.~Auer, N.~Cesa-Bianchi, and P.~Fischer.
\newblock Finite-time analysis of the multiarmed bandit problem.
\newblock {\em Machine Learning}, 47:235 -- 256, 2002.

\bibitem{auer-nonstoch}
P.~Auer, N.~Cesa-Bianchi, Y.~Freund, and R.~Schapire.
\newblock The nonstochastic multi-armed bandit problem.
\newblock {\em SIAM J. Comput.}, 32:48--77, 2002.

\bibitem{awerbuch-kleinberg}
B.~Awerbuch and R.~Kleinberg.
\newblock Adaptive routing with end-to-end feedback: distributed learning and
  geometric approaches.
\newblock {\em STOC}, 2004.

\bibitem{blum-mansour}
A.~Blum and Y.~Mansour.
\newblock From external to internal regret.
\newblock {\em Journal of Machine Learning Research}, pages 1307--1324, 2007.

\bibitem{plg}
N.~Cesa-Bianchi and G.~Lugosi.
\newblock {\em Prediction, Learning and Games}.
\newblock Cambridge University Press, 2006.

\bibitem{normalhedge}
K.~Chaudhuri, Y.~Freund, and D.~Hsu.
\newblock A parameter free hedging algorithm.
\newblock {\em NIPS}, 2009.

\bibitem{cover-binary}
T.~Cover.
\newblock Behaviour of sequential predictors of binary sequences.
\newblock {\em Transactions of the Fourth Prague Conference on Information
  Theory, Statistical Decision Functions, Random Processes}, 1965.

\bibitem{cover-portfolios}
T.~Cover.
\newblock Universal portfolios.
\newblock {\em Mathematical Finance}, 1991.

\bibitem{dani-hayes}
V.~Dani and T.~Hayes.
\newblock Robbing the bandit: Less regret in online geometric optimization
  against an adaptive adversary.
\newblock {\em SODA}, 2006.

\bibitem{kearns-regret}
E.~Even-Dar, M.~Kearns, Y.~Mansour, and J.~Wortman.
\newblock Regret to the best vs. regret to the average.
\newblock {\em Machine Learning}, 72:21--37, 2008.

\bibitem{no-gradient}
A.~Flaxman, A.~Kalai, and H.~B. McMahan.
\newblock Online convex optimization in the bandit setting: Gradient descent
  without a gradient.
\newblock {\em SODA}, 2005.

\bibitem{freund-coin}
Y.~Freund.
\newblock Predicting a binary sequence almost as well as the optimal biased
  coin.
\newblock {\em COLT}, 1996.

\bibitem{freund-schapire-singer-warmuth}
Y.~Freund, R.~E. Schapire, Y.~Singer, and M.~K. Warmuth.
\newblock Using and combining predictors that specialize.
\newblock {\em STOC}, pages 334--343, 1997.

\bibitem{gittins}
J.~C. Gittins.
\newblock {\em Multi-armed Bandit Allocation Indices}.
\newblock John Wiley, 1989.

\bibitem{hazan-kale}
E.~Hazan and S.~Kale.
\newblock Extracting certainty from uncertainty: Regret bounded by variation in
  costs.
\newblock {\em Machine Learning}, 2009.

\bibitem{hazan-kale-invest}
E.~Hazan and S.~Kale.
\newblock On stochastic and worst-case models for investing.
\newblock {\em Advances in Neural Information Processing Systems}, 2009.

\bibitem{hazan-seshadhri}
E.~Hazan and C.~Seshadhri.
\newblock Efficient learning algorithms for changing environments (full version
  available at http://eccc.hpi-web.de/eccc-reports/2007/tr07-088/index.html).
\newblock {\em ICML}, pages 393--400, 2009.

\bibitem{kalai-vempala}
A.~Kalai and S.~Vempala.
\newblock Efficient algorithms for universal portfolios.
\newblock {\em FOCS}, 2000.

\bibitem{kalai-vempala-jcss}
A.~Kalai and S.~Vempala.
\newblock Efficient algorithms for online decision problems.
\newblock {\em JCSS}, 2004.

\bibitem{mab-nips}
Michael Kapralov and Rina Panigrahy.
\newblock Prediction strategies without loss.
\newblock In {\em NIPS}, pages 828--836, 2011.

\bibitem{dichotomies}
R.~Kleinberg and A.~Slivkins.
\newblock Sharp dichotomies for regret minimization in metric spaces.
\newblock {\em SODA}, 2010.

\bibitem{bandits-metric}
R.~Kleinberg, A.~Slivkins, and E.~Upfal.
\newblock Multi-armed bandits in metric spaces.
\newblock {\em STOC}, 2008.

\bibitem{lai-robbins}
T.~L. Lai and Herbert Robbins.
\newblock Asymptotically efficient adaptive allocation rules.
\newblock {\em Advances in Applied Mathematics}, 6:4--22, 1985.

\bibitem{weighted-majority}
N.~Littlestone and M.K. Warmuth.
\newblock The weighted majority algorithm.
\newblock {\em FOCS}, 1989.

\bibitem{tsitsiklis}
J.~Tsitsiklis.
\newblock A short proof of the gittins index theorem.
\newblock {\em Annals of Applied Probability}, 4, 1994.

\bibitem{vovk-game}
V.~Vovk.
\newblock A game of prediction with expert advice.
\newblock {\em Journal of Computer and System Sciences}, 1998.

\bibitem{vovk}
V.~Vovk.
\newblock Derandomizing stochastic prediction strategies.
\newblock {\em Machine Learning}, pages 247–--282, 1999.

\bibitem{zinkevich}
M.~Zinkevich.
\newblock Online convex programming and generalized infinitesimal gradient
  ascent.
\newblock {\em ICML}, 2003.

\end{thebibliography}

\appendix

\section{Properties of $g(x)$} \label{app:A}
In this section we prove  some useful properties of the function $g(x)$.
\begin{equation*}
g(x)=\text{sign}(x) \min\left\{Z \erf\left(\frac{|x|}{4L}\right) e^{\left(\frac{x}{4L}\right)^2}, 1\right\}.
\end{equation*}

First,

{{\bf Fact~\ref{fact-2}} 
One has $g(x)=1$ for $|x|\geq 7L\sqrt{\log(1/Z)}$ as long as $Z\leq 1/e$.
}

\begin{proof}
One has 
\begin{equation*}
Z \erf\left(\frac{x}{4L}\right) e^{\left(\frac{x}{4L}\right)^2}\geq Z\erf(1) e^{(7/4)^2 \log(1/Z)}=\erf(1) e^{((7/4)^2-1) \log(1/Z)}\geq 1
\end{equation*}
when $Z\leq 1/e$.
\end{proof}

{{\bf Fact~\ref{fact-1}}

The function $g(x)=Z \erf \left(\frac{x}{4L}\right) e^{\left(\frac{x}{4L}\right)^2}$ is monotonically increasing and convex for any $Z>0$ for $x\geq 0$.
}

\begin{proof}
One has 
\begin{equation*}
\begin{split}
g'(x)&=\frac{x}{8L^2} g(x)+\frac{Z}{2\sqrt{\pi}L}\\
g''(x)&=\frac{1}{8L^2} g(x)+\frac{x}{8L^2} g'(x)
\end{split}
\end{equation*}
Thus, $g''(x)\geq 0$ for $x\geq 0$.
\end{proof}

Let $U$ be the positive solution of $g(x)=1$, which exists by Fact~\ref{fact-1}. We have $U\leq 7L\sqrt{\log(1/Z)}$ by Fact~\ref{fact-2}.

{{\bf Lemma~\ref{lm:g-prec}} 
For any $|\Delta|\leq 2$ any  $0\leq |x|\leq 2U$ one has 
\begin{equation*}
|g'(x+\Delta)|\leq 1.8 |g'(x)|
\end{equation*}
as long as $L^2\geq 80\log (1/Z)$ and $Z\leq 1/e$.
}

\begin{proof}
First suppose that $x\geq 0$ and $\Delta>0$.
We have 
\begin{equation*}
g'(x)=Z\left(\frac{1}{2\sqrt{\pi}L}+\frac{x}{8L^2}\erf\left(\frac{x}{4L}\right)e^{\frac{x^2}{16L^2}}\right)
\end{equation*}
Thus, 
\begin{equation*}
\begin{split}
g'(x+\Delta)&\leq Z\left(\frac{1}{2\sqrt{\pi T}}+\frac{x+\Delta}{2T}\erf\left(\frac{x+\Delta}{4\sqrt{T}}\right)e^{\frac{(x+\Delta)^2}{16T}}\right)\\
\end{split}
\end{equation*}

We consider two cases:
\begin{enumerate}
\item First suppose that $L/4\leq x\leq 2U\leq 14 L\sqrt{\log (1/Z)}$. Then 
\begin{equation*}
\begin{split}
g'(x+\Delta)&\leq Z\left(\frac{1}{2\sqrt{\pi}L}+\frac{x+\Delta}{8L^2}\erf\left(\frac{x+\Delta}{4L}\right)e^{\frac{(x+\Delta)^2}{16L^2}}\right)\\
\end{split}
\end{equation*}

We bound the relative change in the second term. For the first two factors, we have 
\begin{equation}\label{eq:vue}
\begin{split}
\frac{x+2}{x}&\leq 1+\frac{8}{L}\\
\erf\left(\frac{x+2}{4L}\right)/\erf\left(\frac{x}{4L}\right)&\leq 1/\erf(1)\leq 1.25\\
\end{split}
\end{equation}
Also,
\begin{equation*}
\begin{split}
e^{\frac{(x+\Delta)^2}{16L^2}}&=e^{\frac{x^2}{16L^2}}\cdot e^{\frac{2\Delta x+\Delta^2}{16L^2}}\leq e^{\frac{x^2}{16L^2}}\left(1+\frac{4\Delta x+2\Delta^2}{16L^2}\right)
\end{split}
\end{equation*} 
We have $x\leq 2U\leq 14L\sqrt{\log (1/Z)}$ by the assumptions of the lemma. We have 
\begin{equation*}
\begin{split}
\frac{2\Delta x+\Delta^2}{16L^2}\leq \frac{28 L\sqrt{\log(1/Z)}+4}{16L^2}\leq 2\frac{\sqrt{\log(1/Z)}}{L}+\frac{1}{4L^2}.
\end{split}
\end{equation*} 
Since $L^2\geq  80\log(1/Z)\geq 80$, we can use the inequality $e^x\leq 1+2x, x\in [0, 1]$ to bound the last term:

\begin{equation}\label{eq:ue}
\begin{split}
e^{\frac{(x+\Delta)^2}{16L^2}}&\leq e^{\frac{x^2}{16L^2}}\left(1+\frac{4\Delta x+2\Delta^2}{16L^2}\right)\leq 1.32 e^{\frac{x^2}{16L^2}}.
\end{split}
\end{equation} 

Finally, combining \eqref{eq:vue} and \eqref{eq:ue}, we obtain the desired bound
\begin{equation}
\begin{split}
g'(x+\Delta)\leq 1.8 g(x).
\end{split}
\end{equation} 

\item Otherwise, if $0\leq x\leq L/4$
\begin{equation*}
\begin{split}
g'(x+\Delta)&\leq Z\left(\frac{1}{2\sqrt{\pi} L}+\frac{x+\Delta}{8L^2}\erf\left(\frac{x+\Delta}{4L}\right)e^{\frac{(x+2)^2}{16L^2}}\right)\\
\end{split}
\end{equation*}
Since $L\geq 80$, we have
\begin{equation*}
\begin{split}
\frac{x+\Delta}{8L^2}\erf\left(\frac{x+\Delta}{4L}\right)e^{\frac{(x+2)^2}{16L^2}}\leq \frac{1/4+1/L}{4L}e^{(1+1/(2L))^2}\leq \frac1{5L}.
\end{split}
\end{equation*}
Hence, 
\begin{equation*}
g'(x+\Delta)/g'(x)\leq \frac{\frac{1}{2\sqrt{\pi} L}+\frac1{5L}}{\frac{1}{2\sqrt{\pi} L}}\leq 1+\frac{2\sqrt{\pi}}{5}\leq 1.8.
\end{equation*}
\end{enumerate}
This completes the proof under the assumption that $x\geq 0$ and $\Delta\geq 0$. It remains to note that the general case now follows since $|g'(x)|$ is an even function.
\end{proof}

\end{document}